%% file: main.tex
\pdfoutput=1
\documentclass[12pt,letter]{article}

\usepackage{ifthen} 
\newboolean{pdflatex}
\setboolean{pdflatex}{true} 

\newboolean{articletitles}
\setboolean{articletitles}{true} 

\newboolean{uprightparticles}
\setboolean{uprightparticles}{false} 

\def\paperauthors{Me}
\def\paperasciititle{Substructure of Multiquark Hadrons (White Paper)} 
\def\papertitle{\paperasciititle} 
\def\paperkeywords{{High Energy Physics}, {Hadron Spectroscopy}} 
\def\papercopyright{\quad } 
\def\paperlicenceurl{https://creativecommons.org/licenses/by/4.0/}

\input{preamble} 
\usepackage{longtable} 

\begin{document}

\renewcommand{\thefootnote}{\fnsymbol{footnote}}
\setcounter{footnote}{1}

\input{title-Snowmass-PAPER}


\renewcommand{\thefootnote}{\arabic{footnote}}
\setcounter{footnote}{0}

\tableofcontents
\cleardoublepage


\pagestyle{plain} 
\setcounter{page}{1}
\pagenumbering{arabic}

\input{body}





\clearpage
\addcontentsline{toc}{section}{References}
\bibliographystyle{LHCb}
\bibliography{merged}
 
\end{document}

%% file: preamble.tex
\usepackage[top=1in, bottom=1.25in, left=1in, right=1in]{geometry}

\usepackage{microtype}
\usepackage{lineno}  
\usepackage{xspace} 
\usepackage{caption} 

\usepackage{graphicx}  
\usepackage{color}
\usepackage{colortbl}
\graphicspath{{./figs/}} 

\usepackage{amsmath} 
\usepackage{amssymb}
\usepackage{amsfonts}
\usepackage{upgreek} 

\newcommand*\patchAmsMathEnvironmentForLineno[1]{%
\expandafter\let\csname old#1\expandafter\endcsname\csname #1\endcsname
\expandafter\let\csname oldend#1\expandafter\endcsname\csname
end#1\endcsname
 \renewenvironment{#1}%
   {\linenomath\csname old#1\endcsname}%
   {\csname oldend#1\endcsname\endlinenomath}%
}
\newcommand*\patchBothAmsMathEnvironmentsForLineno[1]{%
  \patchAmsMathEnvironmentForLineno{#1}%
  \patchAmsMathEnvironmentForLineno{#1*}%
}
\AtBeginDocument{%
\patchBothAmsMathEnvironmentsForLineno{equation}%
\patchBothAmsMathEnvironmentsForLineno{align}%
\patchBothAmsMathEnvironmentsForLineno{flalign}%
\patchBothAmsMathEnvironmentsForLineno{alignat}%
\patchBothAmsMathEnvironmentsForLineno{gather}%
\patchBothAmsMathEnvironmentsForLineno{multline}%
\patchBothAmsMathEnvironmentsForLineno{eqnarray}%
}

\usepackage{hyperxmp}

\usepackage[pdftex,
            pdfauthor={\paperauthors},
            pdftitle={\paperasciititle},
            pdfkeywords={\paperkeywords},
            pdfcopyright={Copyright (C) \papercopyright},
            pdflicenseurl={\paperlicenceurl}]{hyperref}

\usepackage[colorinlistoftodos,textsize=scriptsize]{todonotes}

\usepackage[bottom,flushmargin,hang,multiple]{footmisc}

\usepackage[all]{hypcap} 

\input{lhcb-symbols-def} 

\usepackage{cite} 
\usepackage{mciteplus}

%% file: lhcb-symbols-def.tex
\usepackage{xspace} 
\usepackage{upgreek}

\def\MagUp {\mbox{\em Mag\kern -0.05em Up}\xspace}

\ifthenelse{\boolean{uprightparticles}}%
{

 \def\PDelta      {\ensuremath{\Delta}\xspace}                 
 \def\PXi         {\ensuremath{\Xi}\xspace}                 
 \def\PLambda     {\ensuremath{\Lambda}\xspace}                 
 \def\PSigma      {\ensuremath{\Sigma}\xspace}                 
 \def\POmega      {\ensuremath{\Omega}\xspace}                 
 \def\PUpsilon    {\ensuremath{\Upsilon}\xspace}

 \def\PB      {\ensuremath{\mathrm{B}}\xspace}                 
                  
 \def\PD      {\ensuremath{\mathrm{D}}\xspace}

 \def\PK      {\ensuremath{\mathrm{K}}\xspace}

 \def\Pi      {\ensuremath{\mathrm{i}}\xspace}

 \def\Ps      {\ensuremath{\mathrm{s}}\xspace}

 \def\thebaroffset{0.0em}
}
{

 \mathchardef\PDelta="7101
 \mathchardef\PXi="7104
 \mathchardef\PLambda="7103
 \mathchardef\PSigma="7106
 \mathchardef\POmega="710A
 \mathchardef\PUpsilon="7107
                  
 \def\PB      {\ensuremath{B}\xspace}                 
                  
 \def\PD      {\ensuremath{D}\xspace}

 \def\PK      {\ensuremath{K}\xspace}

 \def\Pi      {\ensuremath{i}\xspace}

 \def\Ps      {\ensuremath{s}\xspace}

 \def\thebaroffset{0.18em}
}
\newcommand{\offsetoverline}[2][\thebaroffset]{\kern #1\overline{\kern -#1 #2}}%

\makeatletter
\ifcase \@ptsize \relax
  \newcommand{\miniscule}{\@setfontsize\miniscule{4}{5}}
\or
  \newcommand{\miniscule}{\@setfontsize\miniscule{5}{6}}
\or
  \newcommand{\miniscule}{\@setfontsize\miniscule{5}{6}}
\fi
\makeatother

\DeclareRobustCommand{\optbar}[1]{\shortstack{{\miniscule (\rule[.5ex]{1.25em}{.18mm})}
  \\ [-.7ex] $#1$}}

\def\g      {{\ensuremath{\Pgamma}}\xspace}

\def\squark    {{\ensuremath{\Ps}}\xspace}

\def\KorKbar {\kern \thebaroffset\optbar{\kern -\thebaroffset \PK}{}\xspace}

\def\D       {{\ensuremath{\PD}}\xspace}

\def\DorDbar {\kern \thebaroffset\optbar{\kern -\thebaroffset \PD}\xspace}

\def\Dp      {{\ensuremath{\D^+}}\xspace}
\def\Dm      {{\ensuremath{\D^-}}\xspace}

\def\DpDm    {\ensuremath{\Dp {\kern -0.16em \Dm}}\xspace}

\def\B       {{\ensuremath{\PB}}\xspace}

\def\BorBbar {\kern \thebaroffset\optbar{\kern -\thebaroffset \PB}\xspace}

\def\Bd      {{\ensuremath{\B^0}}\xspace}

\def\BdorBdbar {\kern \thebaroffset\optbar{\kern -\thebaroffset \Bd}\xspace}

\def\Bs      {{\ensuremath{\B^0_\squark}}\xspace}

\def\BsorBsbar {\kern \thebaroffset\optbar{\kern -\thebaroffset \Bs}\xspace}

\def\Y#1S{\ensuremath{\PUpsilon{(#1S)}}\xspace}

\def\LorLbar     {\kern \thebaroffset\optbar{\kern -\thebaroffset \PLambda}\xspace}

\def\to                 {\ensuremath{\rightarrow}\xspace}

\def\AT#1     {\ensuremath{A_{\mathrm{T}}^{#1}}\xspace}           

\def\C#1      {\ensuremath{\mathcal{C}_{#1}}\xspace}                       
\def\Cp#1     {\ensuremath{\mathcal{C}_{#1}^{'}}\xspace}                    
\def\Ceff#1   {\ensuremath{\mathcal{C}_{#1}^{\mathrm{(eff)}}}\xspace}        
\def\Cpeff#1  {\ensuremath{\mathcal{C}_{#1}^{'\mathrm{(eff)}}}\xspace}       
\def\Ope#1    {\ensuremath{\mathcal{O}_{#1}}\xspace}                       
\def\Opep#1   {\ensuremath{\mathcal{O}_{#1}^{'}}\xspace}                    

\newcommand{\aunit}[1]{\ensuremath{\text{\,#1}}}       

\newcommand{\tev}{\aunit{Te\kern -0.1em V}\xspace}
\newcommand{\gev}{\aunit{Ge\kern -0.1em V}\xspace}
\newcommand{\mev}{\aunit{Me\kern -0.1em V}\xspace}
\newcommand{\kev}{\aunit{ke\kern -0.1em V}\xspace}
\newcommand{\ev}{\aunit{e\kern -0.1em V}\xspace}
 
\newcommand{\mevc}{\ensuremath{\aunit{Me\kern -0.1em V\!/}c}\xspace}
\newcommand{\gevc}{\ensuremath{\aunit{Ge\kern -0.1em V\!/}c}\xspace}
\newcommand{\mevcc}{\ensuremath{\aunit{Me\kern -0.1em V\!/}c^2}\xspace}
\newcommand{\gevcc}{\ensuremath{\aunit{Ge\kern -0.1em V\!/}c^2}\xspace}

\def\m    {\aunit{m}\xspace}

\def\nb {\aunit{nb}\xspace}

\def\gsim{{~\raise.15em\hbox{$>$}\kern-.85em
          \lower.35em\hbox{$\sim$}~}\xspace}
\def\lsim{{~\raise.15em\hbox{$<$}\kern-.85em
          \lower.35em\hbox{$\sim$}~}\xspace}

\def\tell1  {TELL1\xspace}
\def\ukl1   {UKL1\xspace}



%% file: title-Snowmass-PAPER.tex

\begin{titlepage}
\pagenumbering{roman}

\vspace*{-1.5cm}
\centerline{\large Snowmass 2021 (DPF Community Planning Exercise) }
\vspace*{0.5cm}
\noindent
\begin{tabular*}{\linewidth}{lc@{\extracolsep{\fill}}r@{\extracolsep{0pt}}}
\vspace*{-1.5cm}\mbox{\!\!\!\includegraphics[width=.14\textwidth]{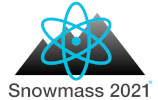}} & &%
\\
 & &  \\   
 &  & \today \\ 
 & & \\
\end{tabular*}

\vspace*{0.2cm}

{\normalfont\bfseries\boldmath\huge
\begin{center}
  \papertitle 
\end{center}
}

\vspace{-0.1cm}

\input{authorship}

\vspace{\fill}

\begin{abstract}
  \noindent
  \input{abstract}
\end{abstract}

\vspace*{2.0cm}


\vspace{\fill}

\vspace*{2mm}

\end{titlepage}


\newpage
\setcounter{page}{2}
\mbox{~}

%% file: authorship.tex

\begin{center}
\small
Nora Brambilla$^{1,2,3}$,
Hua-Xing Chen$^4$,
Angelo Esposito$^5$,
Jacopo Ferretti$^6$,
Anthony Francis$^{7,8,9}$
\hbox{Feng-Kun Guo$^{10,11}$,}
Christoph Hanhart$^{12}$,
Atsushi Hosaka$^{13}$,
Robert L. Jaffe$^{14}$,
\hbox{Marek Karliner$^{15,\dagger}$,}
\hbox{Richard Lebed$^{16}$,}
Randy Lewis$^{17}$,
Luciano Maiani$^{18}$, 
Nilmani Mathur$^{19}$,
\hbox{Ulf-G. Meißner$^{12,20}$,}
Alessandro Pilloni$^{21,22}$,
Antonio Davide Polosa$^{18}$,
Sasa Prelovsek$^{23,24}$,
\hbox{Jean-Marc Richard$^{25}$,}
Ver\'onica Riquer$^{18}$, 
Mitja Rosina$^{23,24}$,
Jonathan L. Rosner$^{26}$,
\hbox{Elena Santopinto$^{27,\ddagger}$,}
\hbox{Eric S. Swanson$^{28}$,}
Adam P. Szczepaniak$^{29,30,31}$, Sachiko Takeuchi$^{32}$,
Makoto Takizawa$^{33}$,
\hbox{Frank Wilczek$^{34,35,36,37,38}$,}
Yasuhiro Yamaguchi$^{39}$,
Bing-Song Zou$^{10,11,40}$.
\end{center}

\smallskip\noindent
{$^\dagger$Corresponding author: \tt marek@tauex.tau.ac.il}
\hfill\break
{$^\ddagger$Corresponding author: \tt elena.santopinto@ge.infn.it}
\begin{center}

\bigskip{\it
\footnotesize
$^{1}$Physik Department, Technische Universität München\\
$^2$Institute for Advanced Study, Technische Universität München\\
$^3$Munich Data Science Institute, Technische Universität München\\
$^4$School of Physics, Southeast University, Nanjing 210094, China\\
$^5$Institute for Advanced Study, Princeton, New Jersey 08540, USA\\
$^6$Physics Dpt., University of Jyv\"askyl\"a, P.O.B. 35 (YFL), 40014 Jyv\"askyl\"a, Finland\\
$^7$Albert Einstein Center, Universität Bern, Sidlerstrasse 5, 3012 Bern, Switzerland\\
$^8$Institute of Physics, National Yang Ming Chiao Tung University, 30010 Hsinchu, Taiwan\\
$^9$Theory Department, CERN, 1201 Geneva, Switzerland\\
$^{10}$CAS Key Laboratory of Theoretical Physics, Institute of Theoretical Physics, Beijing 100190, China\\
$^{11}$School of Physical Sciences, University of Chinese Academy of Sciences, Beijing 100049, China\\
$^{12}$ Institute for Advanced Simulation, Institut für Kernphysik and Jülich Center for Hadron Physics,Forschungszentrum Jülich, D-52425 Jülich, Germany.\\
$^{13}$Research Center for Nuclear Physics (RCNP), Osaka University, Ibaraki, 567-0047, Japan\\
$^{14}$Physics Dpt and LNS, MIT, Cambridge, MA 02139, USA\\
$^{15}$School of Physics and Astronomy, Tel Aviv University, Tel Aviv 69978, Israel\\
$^{16}$Arizona State University, Department of Physics, Tempe, Arizona 85287-1504, USA\\
$^{17}$Dpt of Physics and Astronomy, York University, Toronto, Ontario, M3J 1P3, Canada\\
$^{18}$Dipartimento di Fisica and INFN Sezione di Roma, Sapienza Università di Roma, I-00185 Roma, Italy\\
$^{19}$Department of Theoretical Physics, Tata Institute of Fundamental Physics, Mumbai 400005, India\\
$^{20}$Helmholtz-Institut für Strahlen- und Kernphysik and Bethe Center for Theoretical Physics, Universität Bonn, D-53115 Bonn, Germany\\
$^{21}$Università di Messina, I-98122 Messina, Italy\\
$^{22}$INFN Sezione di Catania, I-95123 Catania, Italy\\
$^{23}$Faculty of Mathematics and Physics, University of Ljubljana\\
$^{24}$Jozef Stefan Institute, 1000 Ljubljana, Slovenia\\
$^{25}$IN2P3-CNRS--UCBL, Universit\'e de Lyon, 69622  Villeurbanne, France\\
$^{26}$Enrico Fermi Institute and Department of Physics, University of Chicago, Chicago, IL 60637, USA\\
$^{27}$INFN Sezione di Genova, 16146 Genova, Italy\\
$^{28}$ Department of Physics and Astronomy,
University of Pittsburgh, Pittsburgh, PA 15260, USA\\
$^{29}$Center for Exploration of Energy and Matter, Indiana University, Bloomington, IN 47403, USA\\
$^{30}$Department of Physics, Indiana University, Bloomington, IN 47405, USA\\
$^{31}$Theory Center, Thomas Jefferson National Accelerator Facility, Newport News, VA 23606, USA\\
$^{32}$Showa Pharmaceutical University, Machida, Tokyo, 194-8543, Japan\\
$^{33}$Japan College of Social Work, Kiyose, Tokyo, 204-8555, Japan\\
$^{34}$Department of Physics, Stockholm University
AlbaNova University Center, 106 91 Stockholm, Sweden\\
$^{35}$ Center for Theoretical Physics, MIT, Cambridge MA 02139, USA\\
$^{36}$ T. D. Lee Institute, Shanghai, China\\
$^{37}$ Wilczek Quantum Center, Department of Physics and Astronomy,\\
Shanghai Jiao Tong University, Shanghai 200240, China\\
$^{38}$Department of Physics and Origins Project, Arizona State
University, Tempe AZ 25287 USA\\
$^{39}$Adv. Science Research Center, Japan Atomic Energy Agency (JAEA), Tokai 319-1195, Japan\\
$^{40}$School of Physics, Central South University, Changsha 410083, China
}
\end{center}

%% file: abstract.tex
%
In recent years there has been a rapidly growing body of experimental evidence for existence of
exotic, {\em multiquark hadrons}, i.e. mesons which contain additional quarks, 
beyond the usual quark-antiquark 
pair and baryons which consist of more than three quarks. In all cases with robust evidence
they contain at least one heavy quark $Q=c\hbox{ or }b$, the majority including two heavy quarks.
Two key theoretical questions have been triggered by these discoveries: 
(a) how are quarks organized inside these multiquark states -- as compact objects with all quarks within 
one confinement volume, interacting via color forces, perhaps with an important role played by diquarks, or as deuteron-like hadronic molecules, bound by light-meson exchange? 
(b) what other multiquark states should we expect? 
The two questions are tightly intertwined.  
Each of the interpretations provides a natural explanation of parts of the data, 
but neither explains all of the data. It is quite possible that both kinds of structures appear in Nature.
It may also be the case that certain states are superpositions of the compact and molecular configurations.
This Whitepaper brings together contributions from many leading practitioners in the field, representing 
a wide spectrum of theoretical interpretations. 
We discuss the importance of future experimental and phenomenological work, which will lead to better understanding 
of multiquark phenomena in QCD. 

%% file: body.tex
\input{mydefinitions}
\input{introduction}

\input{jaffe}
\input{wilczek}
\input{maiani}

\input{rosina}
\input{richard}

\input{lebed}

\input{takizawa}
\input{zou}
\input{chen}
\input{meissner}
\input{guo}
\input{hanhart}

\input{swanson}

\input{pilloni}
\input{prelovsek}

\input{mathur}

\input{francis}
\input{ferretti}

\input{rosner}
\input{hosaka}
\input{brambilla}

\input{yamaguchi}

%% file: introduction.tex
\section{Introduction}
\label{sec:introduction}


Exotic, multiquark states had been predicted \cite{Gell-Mann:1964ewy,Zweig:1964ruk,Zweig:1964jf}
long before the advent of QCD and even longer before the experimental discoveries of the recent years.
In particular, in Ref.~\cite{Gell-Mann:1964ewy}, in which Gell-Mann introduced the idea of quarks, 
with mesons as $\bar q q$ and baryons as $qqq$,
he also pointed out the possibility of $\bar q \bar q q q$ mesons and 
$\bar q q q q q$ baryons. Similar remarks can be found in \cite{Zweig:1964ruk,Zweig:1964jf}.
At that early stage, these had been purely group-theoretical insights,
in modern language, the necessary conditions for forming a color singlet.

In the 1970s Jaffe carried out the first explicit calculations of multiquark states. 
This pioneering effort was based on the dynamical framework of the MIT bag model \cite{Jaffe:1976ig,Jaffe:1976ih}.
Jaffe introduced a specific tetraquark model for light mesons like the $a_0 $ and $f_0$, considering both possibilities, i.e., as composed of four quarks and as diquark-antidiquark systems.

These and other early theoretical efforts triggered many experimental searches, with no clear-cut results.
The situation changed dramatically with the 2003 Belle discovery of $X(3872)$ \cite{Belle:2003nnu}, 
the first unambiguously exotic hadron, subsequently confirmed by many other experiments
\cite{ParticleDataGroup:2020ssz}.

From today's perspective it is clear why the early experimental efforts
did not find clear-cut evidence for multiquark states. 
The point is that, in order for a multiquark state to be clearly identifiable, it is not enough  
to form a multiquark color-singlet.
Such a state needs to be narrow enough to stand out on top of the experimental background, and has to have
distinct decay modes which cannot be explained by decay of a conventional hadron.
Multiquark states containing only light quarks typically have many open 
decay channels, with a large phase space, so they tend to be wide. 
Moreover, they share these decay channels with excited states of conventional
hadrons and mix with them, so they are extremely difficult to pin down.

Multiquark states with heavy quarks are very different.
This is where QCD dynamics enters. To paraphrase Orwell:   {\em all quarks are equal, 
but the heavy quarks are more equal then others}. Simply put, all quarks couple in the same way to gluons.
Moreover,
for light quarks the ratio $\,m_q/\Lambda_{QCD}$ is small, and can be neglected in zeroth order. 
But for heavy quarks ($c$ or $b$) the ratio $m_Q/\Lambda_{QCD}\gg 1$ is an additional relevant parameter 
which dramatically affects the dynamics and the experimental situation, creating narrow multiquark states
which stand out. These states were not seen in the early searches simply because the relevant production 
cross sections are very small. They became accessible only with the advent of the huge luminosity 
provided by the $B$ factories.

\vrule width 0pt height 2.8ex
To reiterate, multiquark hadrons containing heavy quarks ($c$ or $b$) bypass the obstacles present in the light-only sector:
\begin{itemize}
\item[--]
First, the large mass of the heavy quarks greatly reduces their kinetic energy, making it easier
for them to form multiquark clusters with the light quarks.
\item[--]
Second, the presence of both 
heavy and light quarks makes the initial state unambiguous, e.g., from the decay $Z_b^+ \to \Upsilon \pi^+$ 
\cite{Belle:2011aa} it is
obvious that its quark content is $\bar b b u \bar d$, and from the decay $Z_c(3900)^+ \to J/\psi \pi^+$
\cite{BESIII:2013ris,Belle:2013yex} 
it is clear that that the corresponding quark content is $\bar c c u \bar d$, while 
$\,T_{cc}^+ \to D^0 D^0\pi^+$ sharply peaking at the $D^{*+} D^0$ threshold 
definitely identifies its quark content as $cc \bar u \bar d$ \cite{LHCb:2021vvq,LHCb:2021auc}.
\item[--]
Third, the internal 
structure of many such heavy-light systems likely provides a natural mechanism resulting in a narrow width, 
making them very conspicuous.
\item[--]
Fourth, the attraction between two heavy quarks
scales like $\,\alpha_s^2 m_Q$, growing approximately linearly with the heavy quark mass. 
At least in one case, the $b b \bar u \bar d$ tetraquark, this results in an expected state which is below two-meson threshold, so {\em it is stable under the strong interactions} and can only decay weakly. 

\end{itemize}

%
Indeed, in recent years there has been a rapidly growing body of experimental evidence for existence of
exotic, {\em multiquark hadrons}, i.e., mesons which contain additional quarks, 
beyond the usual quark-antiquark 
pair, and baryons which consist of more than three quarks. In all cases with robust evidence,
they contain at least one heavy quark $Q=c\hbox{ or }b$, the majority including two heavy quarks.
Two key theoretical questions have been triggered by these discoveries: 
\begin{itemize}
\item[(a)]
How are quarks organized inside these multiquark states -- as compact objects with all quarks within 
one confinement volume, perhaps with an important role played by diquarks, or as deuteron-like 
hadronic molecules? 
\item[(b)]
What other multiquark states should we expect? 
\end{itemize}
The two questions are tightly intertwined.  
Each of the interpretations provides a natural explanation of parts of the data, 
but neither explains all of the data. It is quite possible that both kinds of structures appear in Nature.
It may also be the case that certain states are superpositions of the compact and molecular configurations.

This white paper brings together contributions from many leading practitioners in the field, representing 
a wide spectrum of theoretical interpretations. 
We discuss the importance of future experimental and phenomenological work, which will lead to a better understanding 
of multiquark phenomena in QCD. 

 \vfill\eject


%% file: jaffe.tex
\vfill\eject
\color{black}
\section{Pioneering papers on multiquark hadrons}
\centerline{\bf \large Robert L. Jaffe}
\centerline{Physics Dpt and LNS, MIT, Cambridge, MA 02139, USA}
\centerline{\tt jaffe@mit.edu}
\hfill\break

\begin{itemize}
\item
First papers explicitly describing multiquark hadrons, including the 
conjecture that the light scalar mesons are $qq\bar q\bar q$ states
\cite{Jaffe:1976ig,Jaffe:1976ih}.

\item Paper where the existence of (an exotic) dihyperon bound state is
conjectured.  Recent lattice calculations suggest that this is probably
a virtual state somewhat like the di-neutron \cite{Jaffe:1976yi}.

\item
Diquarks and exotic spectroscopy \cite{Jaffe:2003sg}.


\end{itemize}

%% file: wilczek.tex
\vfill\eject
﻿

\section{Hadron Systematics and Emergent Diquarks}
\centerline{\bf\large Frank Wilczek$^{1,2,3,4,5}$}
\centerline{$^1$Department of Physics, Stockholm University, 
AlbaNova University Center, 106 91 Stockholm, Sweden}
\centerline{$^2$Center for Theoretical Physics, MIT, Cambridge MA 02139, USA}
\centerline{$^3$T. D. Lee Institute, Shanghai, China}
\centerline{$^4$Wilczek Quantum Center, Department of Physics and
Astronomy,}
\centerline{Shanghai Jiao Tong University, Shanghai 200240, China}
\centerline{$^5$Department of Physics and Origins Project, Arizona State
University, Tempe AZ 25287 USA}
\centerline{\tt wilczek@mit.edu}
\hfill\break

The attraction between quarks is a fundamental aspect of QCD, and it
is plausible that several of the most profound aspects of low-energy QCD
dynamics are connected to diquark correlations, such as the similarity of
mesons and baryons, color superconductivity at high density, hyperfine
splittings, the $\Delta I = \frac 1 2$ rule, and some striking features
of structure and fragmentation functions. These issues were proposed in
\cite{Wilczek:2004im}. Approximate mass differences for diquarks with
different quantum numbers were given too, as well as a mass-loaded
generalization of the Chew-Frautschi formula \cite{Wilczek:2004im}.
Diquarks and exotic spectroscopy, and in particular the light
pentaquark, were studied in \cite{Jaffe:2003sg}. The importance of diquark
correlations has been stressed in Ref.~\cite{Selem:2006nd}.  A variety of
theoretical and phenomenological indications for the probable importance
of powerful diquark correlations in hadronic physics were discussed in
detail. Moreover, it was demonstrated that the bulk of light-hadron
spectroscopy could be organized using three simple hypotheses: the
Regge-Chew-Frautschi mass formulae, the feebleness of spin-orbit
forces, and energetic distinctions among a few different diquark
configurations \cite {Selem:2006nd}. These hypotheses were implemented
in a semi-classical model of color flux tubes, extrapolated down from
large orbital angular momentum $L$.  Effects of diquark correlations in
observed patterns of baryon decays were discussed too \cite{Selem:2006nd}.

%% file: maiani.tex
\vfill\eject

\section{Exotics as compact tetraquarks}
\label{sec:X(3872)}


\centerline{\bf\large Angelo Esposito$^{1,*}$, Luciano Maiani$^{2,\dagger}$, 
Antonio Davide Polosa$^{2,\S}$
}
\centerline{\bf\large and Ver\'onica Riquer$^{2,\ddagger}$ 
}

\vspace{0.5em}

\centerline{$^1$School of Natural Sciences, Institute for Advanced Study,}
\centerline{1 Einstein Drive, Princeton, NJ 08540, USA}
\centerline{$^2$Dipartimento di Fisica and INFN Sezione di Roma, }
\centerline{Sapienza Università di Roma, Piazzale Aldo Moro 5, I-00185 Roma, Italy}

\vspace{0.5em}

\centerline{\tt $^*$angeloesposito@ias.edu, $^\dagger$luciano.maiani@cern.ch,  $^\S$antoniodavide.polosa@uniroma1.it, }
\centerline{\tt $^\ddagger$veronica.riquer@cern.ch}
\hfill\break



Starting from 2016, new kinds of exotic hadrons have been discovered, in particular        
$J/\Psi \phi$ resonances, di-$J/\Psi$ resonances,  as well as open strangeness states, i.e. the $Z_{cs}(3082)$ and $Z_{cs}(4003)$.
For these particles, one-pion, long range, exchange forces are not present, to produce the hadron molecule made by color singlet mesons. So called molecular models have to stand on the existence of completely {\it ad hoc } phenomenological forces and undetermined parameters, which are difficult to justify. In  addition, the new states are not necessarily {\it just on threshold}.
On the other hand, multiquark states bound  in color singlets by QCD~\cite{Maiani:2004vq,Maiani:2014aja,Maiani:2015vwa,Maiani:2017kyi} can very  well explain the New Exotics~\cite{Maiani:2016wlq,Becchi:2020uvq}.
 
\begin{figure}[htbp]
   \centering
   \includegraphics[width=0.6 \linewidth]{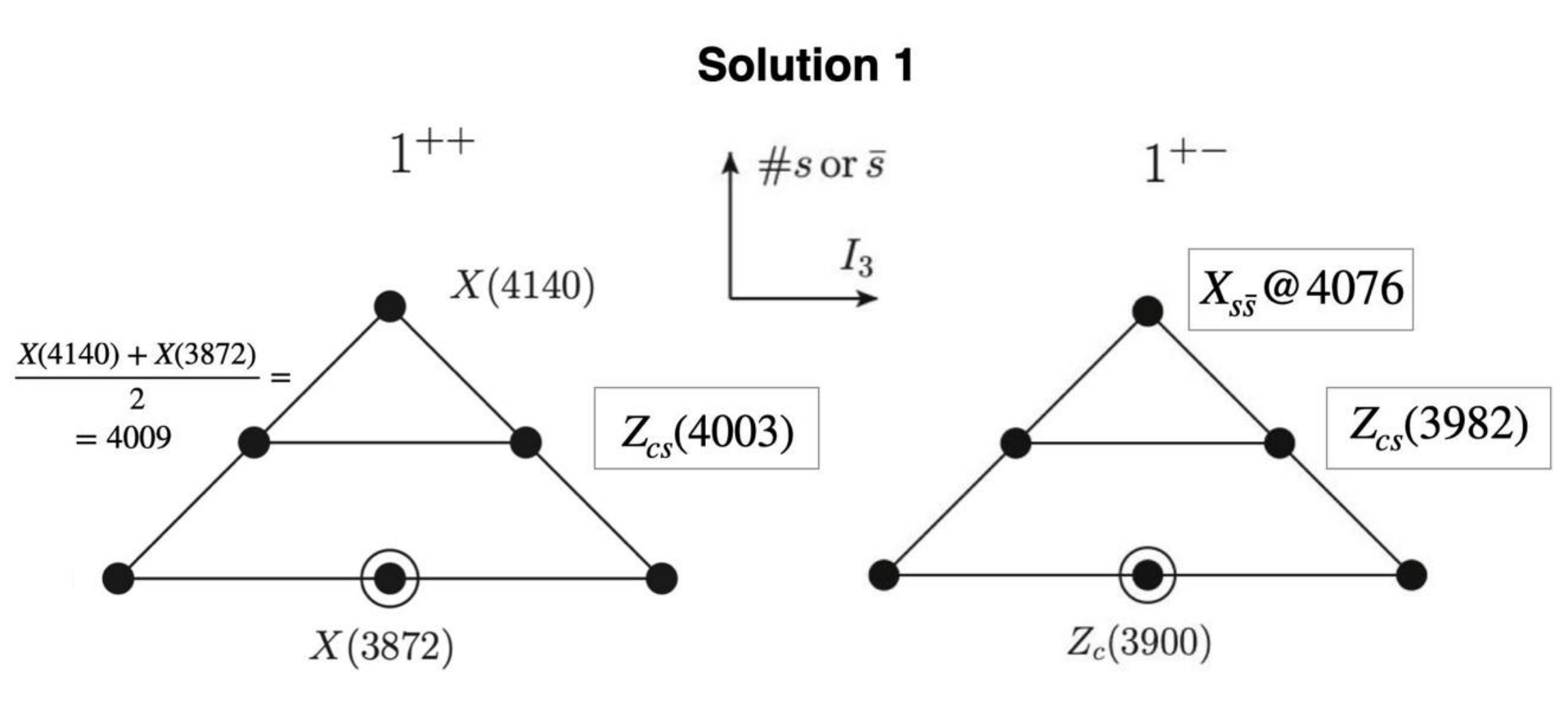}
   \caption{\footnotesize{Possible tetraquark nonets, including the prediction for the yet unobserved $X_{s\bar s}$ state. Taken from~\cite{Maiani:2021tri}.}} 
\label{mass}
\end{figure}

A firm prediction is that hidden charm compact  tetraquarks must form  complete multiplets of flavor $SU(3)$,
with mass differences determined by the strange quark mass difference: $m_s-m_u=120-150$~ MeV. Indeed, with $Z_{cs}(3082)$ and $Z_{cs}(4003)$ we can almost fill two tetraquark nonets with the expected scale of mass differences~\cite{Maiani:2021tri}, see~Fig.~\ref{mass}. 
\vrule width 0pt depth 6ex 

A long standing issue is also how to discriminate between hadronic molecules and compact tetraquarks, and in particular their production and evolution in prompt hadronic collisions. Most of the progress in this direction has been made for the $X(3872)$. In particular, a comparison between the prompt production cross section of the latter with that of other \emph{bona fide} hadronic molecules, shows that the $X(3872)$ is produced much more copiously that one would expect from a loosely bound state~\cite{Esposito:2015fsa}. It has also been shown that its production and evolution in high-multiplicity collisions is well described assuming a behavior not too dissimilar from a charmonium. If the $X(3872)$ were a molecule, its evolution in a dense QCD medium should be qualitatively at odds with data, as shown in~\cite{Esposito:2020ywk}.

Some more progress in discriminating the two models has also been made recently, thanks to some high precision data by LHCb about the $X(3872)$ and $T_{cc}^+(3875)$. Let us take the $X(3872)$ as an example and let us consider the $D^*\bar D$ scattering amplitude.
At low energies, the latter can be expanded as
\begin{align}
f =\frac{1}{ k \cot \delta(k) - i k}=\frac{1}{-\kappa_0 + \frac{1}{2} r_0 k^2+ \dots - i k }\,,
\end{align}
where $\kappa_0$ is the inverse scattering length and $r_0$ the effective range. The latter is a model independent indicator of the microscopic nature of the state. In particular, following Weinberg's criterion, if $|r_0| \lesssim m_\pi^{-1}$, the state is compatible with a composite nature, while if  $r_0<0$ and substantially larger than $m_\pi^{-1}$, it can only be an interacting compact state, see~ \cite{Esposito:2021vhu} for details.

\begin{figure}[htbp]
   \centering
   \includegraphics[width=0.7 \linewidth]{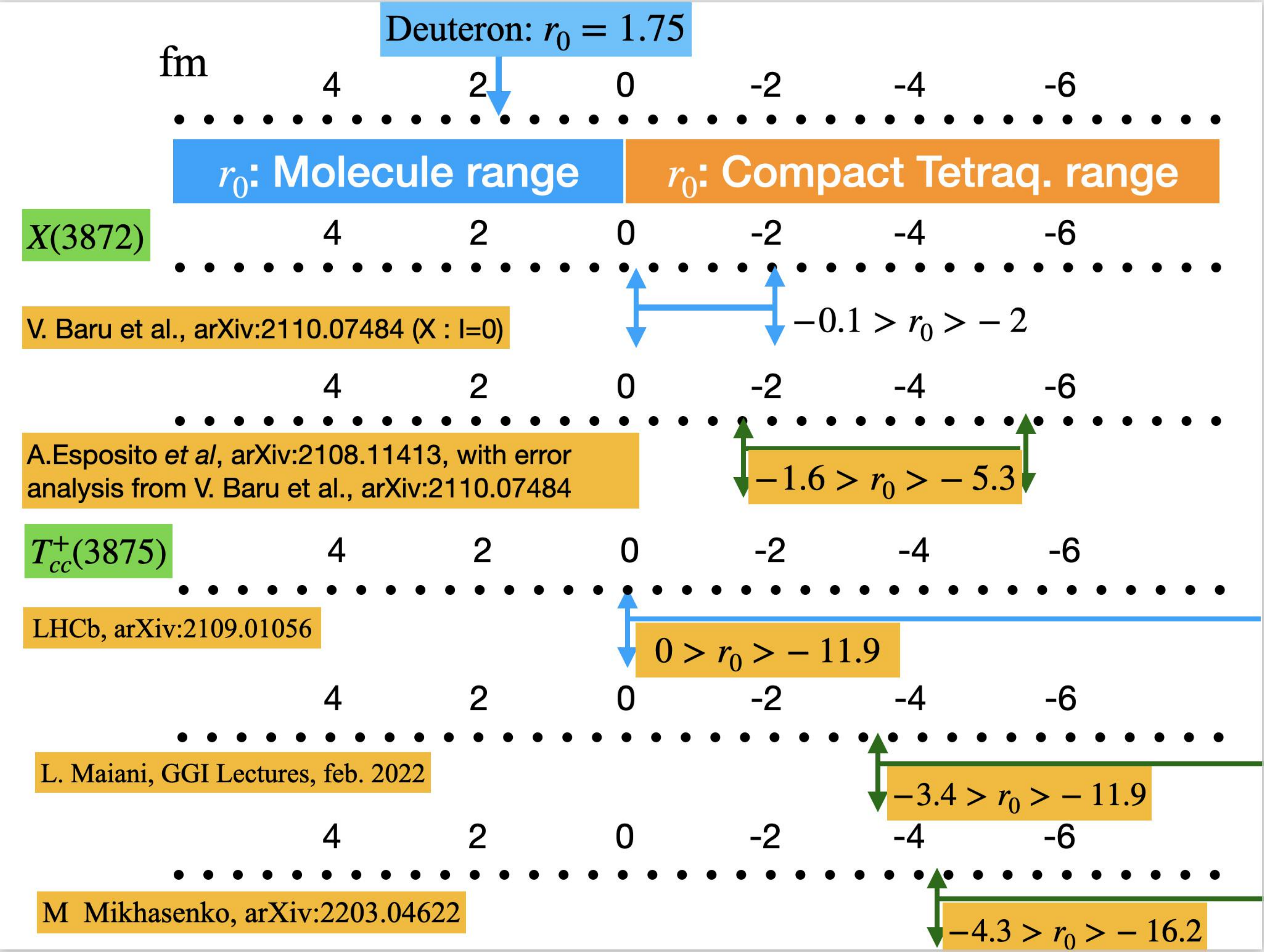}
   \caption{\footnotesize{Schematic summary of the present determinations of $r_0$. }} 
\label{rad}
\end{figure}

Recent LHCb analyses allow to extract information about $r_0$ for both the $X(3872)$ and the $T_{cc}^+(3875)$. The current scenarios are reported in Figure~\ref{rad}. A new  analysis by the Valencia group claims $r_0\simeq +1$~fm for $T_{cc}^+$.
As one can see, there is still no consensus on the matter, but this seems like a promising road to further pursue.


%% file: rosina.tex
\vfill\eject

\section{A short outline of arguments  for \boldmath $T_{cc}^+$ }
\centerline{\bf\large Mitja Rosina}
\centerline{Faculty of Math and Physics, University of Ljubljana}
\centerline{J. Stefan Institute, Ljubljana, Slovenia}
\centerline{\tt mitja.rosina@ijs.si}
\hfill\break

%
%
%
\subsection*{Is the tetraquark \boldmath $T_{cc}^+ \equiv  cc\bar u\bar d$ \,bound?}

The $T_{cc}^+$ tetraquark ( $\equiv$ $DD^*$ dimeson) 
presents a delicate test of
our understanding of quark models. A heavier tetraquark, such as
$T_{bb}^-$,
would be strongly bound against the $B + B^*$ decay, because of the small
kinetic energy, while a lighter tetraquark such as $T_{ss}^-$  is unbound. 
The $T_{cc}^+$ is just on the edge --  most calculations did not  get binding, while ours
did \cite{JancRosina1}.
With the OGE + linear confining  interaction we
obtained  -0.6 MeV (-2.7 MeV) binding energy for the Bhaduri (Grenoble)
parameters. The recent LHCb experiment confirmed the range of our
predictions  (some uncertainty is due to our taking  average of the thresholds
D*$^+$D$^0$ and D*$^0$D$^+$  which differ by 1.4 MeV).

\hfill\break
The main arguments for binding are as follows. 
\begin{itemize}
\item [(i)]
Unlike the protons in the hydrogen molecule, the two $c$ quarks
attract each other at short distances, since the quantum numbers
$I=0$, $J=1^+$ allow them to recouple to the attractive  color 
triplet state.
\item[(ii)]
A very rich four-body space is needed  to get  binding. Neither
the atom-like configurations of the $cc$-diquark  with light antiquarks
around, nor the $D D^*$ molecular configurations alone are sufficient. 
\item[(iii)]
The results are not very sensitive to the model parameters, provided they
reproduce several heavy baryons and mesons whose wavefunctions resemble
important components of the tetraquark wavefunction.
\end{itemize}

\subsection*{The synthesis of the \boldmath $T_{cc}^+$ tetraquark}

The LHCb experiment has identified $T_{cc}^+$ via its decay, 
as a D$^0$D$^0 \pi^+$ resonance.
For future experiments, however, it would be useful to understand also
the production mechanism. In \cite{JancRosina2} we have proposed a 4-step mechanism:  
\begin{enumerate}
\item
production of two $c \bar c$ pairs via double two-gluon fusion:
\hfill\break
$(g {+} g) + (g {+} g)  \to  (c {+} \bar c) + (c {+} \bar c)$; 
\item
the two $c$-s  fly close enough in phase space to bind:
\hfill\break
$(c + c) \to  cc$ (colour antisymmetric diquark); 
\item the $cc$ diquark gets dressed: \ 
$cc \to  ccu$ \,or\, $cc \bar u \bar d$; 
\item
dimeson forms: $cc \bar u \bar d \to  D D^*$. 
\end{enumerate}
The above mechanism leads to
a sufficient number of tetraquarks, and in fact,  the LHCb experiment
has seen them. Further analysis is needed.

\subsection*{The width of the \boldmath $T_{cc}^+$ tetraquark}

The width of  $T_{cc}^+$ is related to the widths of its constituent 
$D^{*0}$ and $D^{*+}$.
Unfortunately, reliable values of the $D^{*0}$ and $T_{cc}^+$
widths are not yet known and at least theoretical estimates would be
welcome. We have estimated the relation using a toy model with complex
poles \cite{Rosina2021}: 

$$\begin{pmatrix}
m1 - i  \Gamma_1 /2 & k \\
    k               & m2  - i \Gamma_2 /2
\end{pmatrix}
\begin{pmatrix}
x \\ y
\end{pmatrix}
=
(m - i \Gamma /2) 
\begin{pmatrix}
x \\ y
\end{pmatrix}
\,. 
$$
Here $m1$, $m2$ and $m$ are the experimental $D^{*+} D^0$ and $D^{*0} D^+$
thresholds and tetraquark masses, respectively.
$\Gamma_1$ = 83.4 keV and  $k$ is a  coupling fitted to reproduce these
values. Then the relation between the guesses for the  $\Gamma_2$ and
$\Gamma$ widths follows. 
For example, $\Gamma_2 = 55$ keV $\to \Gamma =
59$ keV, or  $\Gamma_2 = 2400$ keV $\to \Gamma = 410$ keV.

An interesting effect is due to the charge splitting of the $D^{*+} D^0$
and $D^{*0} D^+$ thresholds, therefore $T_{cc}^+$  will not have a pure 
isoscalar coupling 
$T_{cc}^+ = (D^{*+} D^0 - D^{*0} D^+)/\sqrt{2}$\, and the actual
composition will be seen in the branching ratios. 


%% file: richard.tex
\vfill\eject
\section{Contribution of Jean-Marc~Richard}
\centerline{Universit\'e de Lyon, Institut de Physique des 2 Infinis de Lyon,
IN2P3-CNRS--UCBL}
\centerline{4 rue Enrico Fermi, 69622  Villeurbanne, France}
\centerline{\tt j-m.richard@ipnl.in2p3.fr
\vrule width 0pt depth 3.5ex}
%
%
\subsection*{Chromoelectric binding of tetraquarks}
\label{se:chromo-e}
The possibility of multiquark resonances and bound states due to the \emph{chromomagnetic} interaction has been stressed during the 70s, and has received a lot of attention. 

In \cite{Ader:1981db}, a \emph{chromoelectric} mechanism has been revealed: a system $QQ\bar q\bar q$ interacting via a spin-independent interaction $v(r)$, e.g., of Coulomb-plus-linear type, with color factors corresponding to the exchange of a color-octet, becomes stable if the quark-to-antiquark mass ratio $M/m$ is large enough, i.e., $QQ\bar q\bar q< Q\bar q+Q\bar q$.


In practice, within current quark models, a pure chromo-electric binding is elusive, since it requires a very large $M/m$. However, for $cc\bar u\bar d$ and the analogs with $c\to b$, in a state $J^P=1^+$, the $\bar u\bar d$ pair is mainly in a spin 0 and isospin 0 state, and thus receives an attractive chromomagnetic contribution that has no counterpart in the threshold. Hence the binding of $QQ\bar u\bar d$ is due to a cooperative effect of the chromoelectric interaction  between the two heavy quarks and  chromomagnetic interaction between the two light antiquarks. 
\subsection*{Other configurations}
Contrary to a current belief, the simple quark model, if treated correctly, does \emph{not} predict a proliferation of bound states. For instance, no $cc\bar c\bar c$ or $bb\bar b\bar b$bound state is found below the threshold made of two quarkonia,
%
 and similarly, no $QQqqqq$ hexaquark is found bound below the lowest of the $QQq+qqq$ and $Qqq+Qqq$ thresholds
nor any  all-heavy hexaquark such as $cccbbb$ below the lowest threshold~\cite{Richard:2020zxb}. 

In the pentaquark sector, the studies have been focused mainly on the states found by LHCb or analogs, for the detection of which a $J/\psi$ trigger is crucial.  A thorough survey of pentaquark configurations within a model combining  chromoelectric and chromomagnetic terms indicates, however, the possibility of bound states that would require other triggers. See \cite{Richard:2017una}. This indicates that while LHCb has been essential in the renewal of the physics of heavy exotics, new triggers and new analyses are required for the continuation of this program.

%
\subsection*{Resonances}
The simplest tools developed for the quark model are  suited for
bound-state calculations, and cannot be applied as such to describe
resonances in the continuum. Resonances, that correspond in constituent
models to peaks in the density of states, can be reached with dedicated
techniques that have been elaborated in atomic and nuclear physics. 
%
\subsection*{String dynamics}
\begin{figure}[ht!]
 \centering
 \includegraphics[width=.2\textwidth]{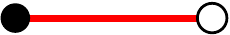} \qquad\includegraphics[width=.15\textwidth]{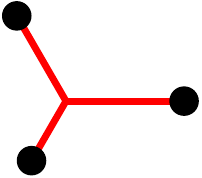}
 \qquad\includegraphics[width=.2\textwidth]{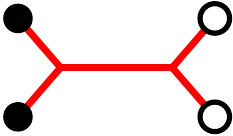}\\[10pt]
 \includegraphics[width=.2\textwidth]{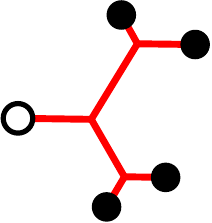} \qquad\includegraphics[width=.2\textwidth]{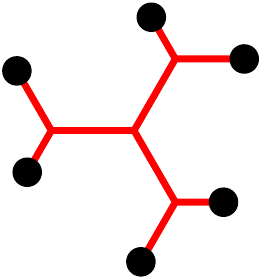}
 \caption{String picture of confinement, for mesons, baryons, tetraquarks, pentaquarks and hexaquarks.}
 \label{FigJMR}
\end{figure}

A variant of the simple quark model describes the confinement as the minimal length (times the string tension) of the flux tubes linking the quarks and antiquarks. See Fig.~\ref{FigJMR}. It gives an interaction which is more attractive than the usual pairwise ansatz for the linear potential, provided it is not suppressed by the constraints of antisymmetrization. Thus a minute comparison of exotics involving various combinations of flavor could probe the dynamics of confinement. 

\strut  









%

%% file: lebed.tex
\vfill\eject

\color{black}
\def\de{\delta^{\vphantom{1}}}
\def\bde{{\bar\delta}}
\def\bt{{\bar\theta}}

\section{Tetraquarks as dynamical  diquarks-antidiquarks
}
\centerline{\bf\large Richard Lebed}
\centerline{Arizona State University, Department of Physics
Tempe, Arizona 85287-1504, USA}
\centerline{\tt richard.lebed@asu.edu}
\hfill\break
Lebed's most important work on exotic hadrons uses diquarks in their
attractive color-triplet channel as hadron constituents.  The primary
difference of his work from other diquark models is the
introduction of a dynamical effect to ensure that the diquarks formed
do not instantaneously rearrange into di-meson pairs.  In the
dynamical diquark {\em picture\/} for tetraquarks introduced in
Ref.~\cite{Brodsky:2014xia}, the relative momentum between the
initial quark (or antiquark) pair that coalesces into a diquark
quasiparticle is much smaller than that between either quark and
either antiquark.  The configuration develops in time into a
spatially separated diquark-antidiquark ($\de$-$\bde$) pair, and
since $\de$ and $\bde$ are not color singlets, they are bound by
confinement, with their large initial relative kinetic energy
eventually converted into the potential energy of a static color flux
tube.  Originally, the picture was motivated by the observation that
the state $Z_c(4430)$ does not lie close to a natural di-hadron
threshold, but also dominantly decays to $\psi(2S)$ rather than to
$J/\psi$, despite much greater phase space in the latter channel,
thus suggesting a large spatial separation between the $c\bar c$ pair
in $Z_c(4430)$.  In order to obtain diquarks sufficiently small to be
identifiable in the hadron substructure, each quasiparticle must
contain a heavy quark $Q$.  The picture can also be generalized to
encompass pentaquarks by noting that the color-triplet attraction can
be extended to build states not just from diquarks
[$\de \! \equiv \! (Qq)_{\bar {\bf 3}}$] but also
{\it triquarks}~[$\bt \! \equiv \! (\bar Q_{\bar {\bf 3}}
(q_1 q_2)_{\bar {\bf 3}})_{\bf 3}$].

This picture was developed into the dynamical diquark
{\em model\/}~\cite{Lebed:2017min} by noting that the color flux
tube, connecting color-triplet and -antitriplet sources, is exactly
the same as the one calculated in lattice simulations for quarkonium
and its hybrids.  Then, using the language of the Born-Oppenheimer
(BO) approximation, the spectrum is developed by combining the (now)
static heavy quasiparticles with the quantum numbers of the light
glue-field degrees of freedom.  For example, the ground-state
tetraquark multiplet $\Sigma^+_g(1S)$ consists of 6 isosinglets and 6
isotriplets, all with positive parity.  Each BO potential
($\Sigma^+_g$, $\Delta^-_u$, {\it etc.}) can be fed into a
Schr\"{o}dinger equation to obtain predictions for the mass
eigenvalues and wave functions of the states, requiring as input only
a choice for the diquark mass $m_\de$.  The first numerical results
of the model (BO multiplet-average masses) required developing a
coupled-channel Schr\"{o}dinger equation solver, since some of
the BO potentials become degenerate in the limit of zero
quasiparticle separation.  Taking $X(3872)$ to be a state in the
multiplet $\Sigma^+_g(1S)$, then the masses for $1^{--}$ states like
$Y(4220)$ are found to coincide with the calculated mass of
$\Sigma^+_g(1P)$ states, and $Z_c(4430)$ fits extremely
well with the calculated mass value for $\Sigma^+_g(2S)$.

\begin{figure}
\centering
\includegraphics[width=0.93\columnwidth]{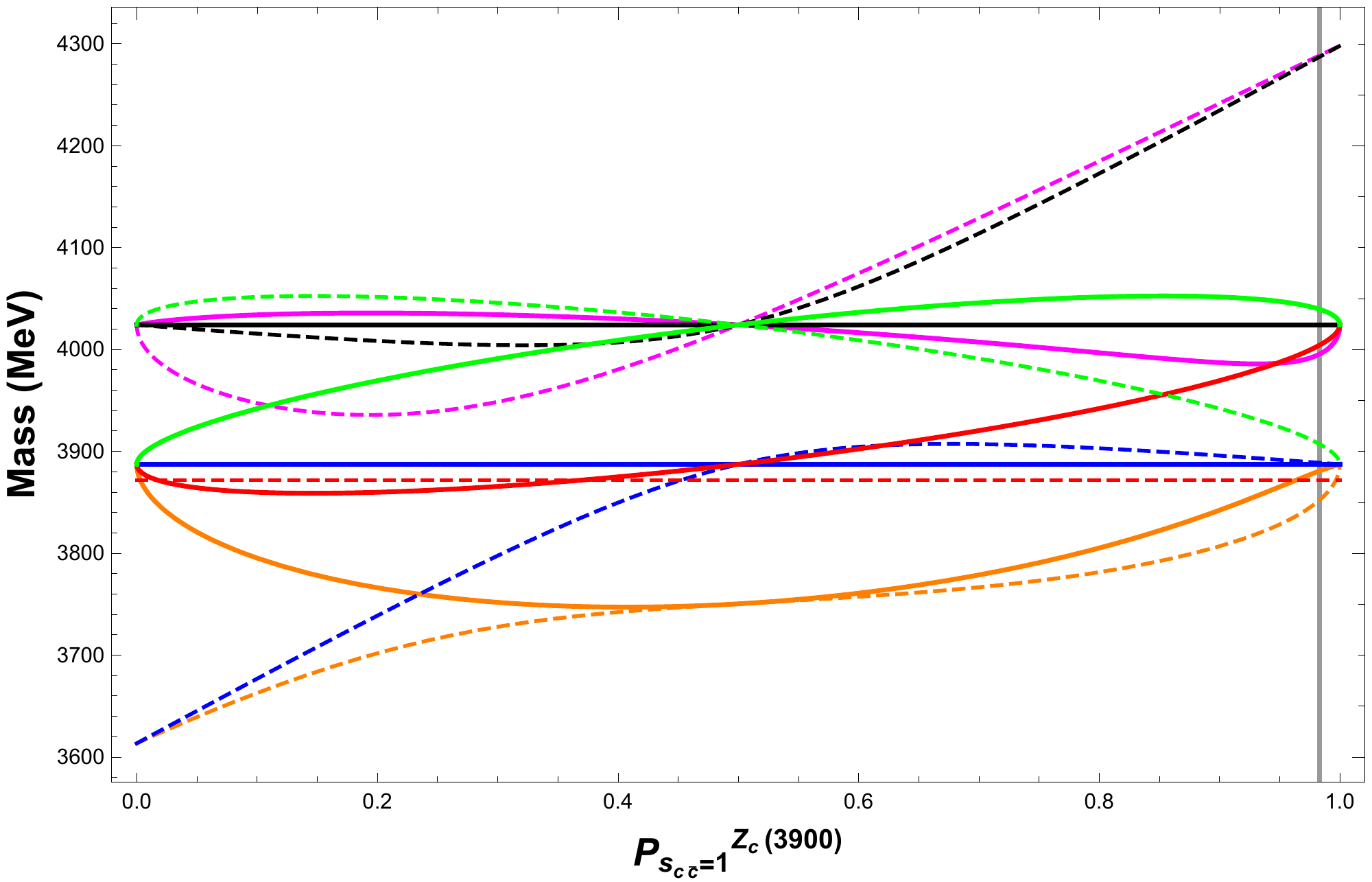}
\caption{Prediction of the 12 isomultiplet masses
(in MeV) of the $\Sigma^+_g(1S)$ hidden-charm multiplet as
functions of the heavy-quark $s_{c\bar c} \! = \! 1$ spin-content
parameter $P$ of $Z_c(3900)$.  Solid (dashed) lines indicate
$I \! = \! 1$ ($I \! = \! 0$) states.  Magenta and gold lines are
$J^{PC} \! = \! 0^{++}$, red lines are $1^{++}$, black and blue
lines are $1^{+-}$, and green lines are $2^{++}$.  The grey vertical
band is the value $P \! \simeq \! 0.983$ obtained from the
3-parameter Hamiltonian of Ref.~\cite{Giron:2019cfc}.  Figure from
J.F.~Giron, PhD thesis (2021).
\label{fig:ccqqJesse}
}
\end{figure}

Fine structure was introduced in the model in Ref.~\cite{Giron:2019cfc} by
means of a Hamiltonian with only two symmetry-breaking parameters:
a spin-spin coupling between the quarks within each diquark, and a
spin-isospin coupling (modeled upon the nucleon-pion-exchange
potential) between the $u$ or $d$ quarks within separate
quasiparticles.  The isosinglet $X(3872)$ emerges naturally as the
lightest clearly identifiable $c\bar c q\bar q^\prime$ state over
broad ranges of the possible parameter space.  A key observable in
these studies is the heavy-quark spin content $s_{Q\bar Q}$ of the
state, evident in quarkonium decays such as in the strong preference
for $Z_c(3900) \! \to \! J/\psi$ and $Z_c(4020) \! \to \! h_c$.  The
3-parameter Hamiltonian predicts this parameter $P$, the
$s_{c\bar c} \! = \! 1$ content for $Z_c(3900)$, to be over 98\%.  As
seen in Fig.~\ref{fig:ccqqJesse}, the mass ordering of the
hidden-charm tetraquarks depends strongly upon $P$, and correlates
with the lightness of $X(3872)$.  Most significantly, masses for all
states in the spectrum are thusly computed.

A number of other results are worth mentioning: (1) The
multiplet $\Sigma^+_g(1P)$ [containing $Y(4220)$] has been studied in
the model, and the $0^{--}$ state $R_{c0}(4240)$ was found to belong
naturally to the multiplet. (2) The application to
$b\bar b q\bar q^\prime$ and $c\bar c s\bar s$ states has been
carried out, using the assignment that $X(3915)$ is the lightest
$c\bar c s\bar s$ state.  Here, one prediction is that $X(4274)$ is
{\em not\/} a tetraquark state, much more likely being the
conventional charmonium state $\chi_{c1}(3P)$. (3) $c\bar c c\bar c$
states have been studied, with the result that $X(6900)$ is a
$\Sigma^+_g(2S)$ state, and observed structure near 7200~MeV is due
to the breaking of the flux tube at the $\Xi_{cc} {\bar \Xi}_{cc}$
threshold.  (4) $c\bar c q\bar s$ states have been studied, and the
large $Z_{cs}(4220)$-$Z_{cs}(4000)$ splitting is found to be
explained by mixing of SU(3)$_{\rm flavor}$ multiplets, like what
occurs for the light hadrons $K_{1A}, K_{1B}$.  (5) Lastly,
hidden-charm pentaquarks have been studied, and the closely spaced
pairs $P_c(4337)$-$P_c(4312)$ and $P_c(4457)$-$P_c(4440)$ arise as
${\frac 1 2}^+ \!$-${\frac 3 2}^+$ pairs in the multiplet
$\Sigma^+(1P)$.  In all cases, all of the missing-state masses in
each multiplet are predicted.

%% file: takizawa.tex
\vfill\eject
{\color{black}

\section{\boldmath $X(3872)$ in hybrid charmonium-$D \bar D^*$ model, 
\hfill\break
its decays and isovector pentaquarks}
\centerline{\bf\large Sachiko Takeuchi$^{1*}$ and Makoto Takizawa$^{2\dagger}$}
\centerline{$^1$Showa Pharmaceutical University, Machida, Tokyo, 194-8543, Japan}
\centerline{$^2$Japan College of Social Work, Kiyose, Tokyo, 204-8555, Japan}
\centerline{\tt $^*$takizawa@ac.shoyaku.ac.jp\qquad $^\dagger$s.takeuchi@jcsw.ac.jp}
\hfill\break


We have studied the structure of $X(3872)$ in the charmonium and
$D^0 \bar D^{\ast 0}$ and $D^+ \bar D^{\ast -}$ hybrid model in
Ref.~\cite{Takizawa:2012hy}. The strengths of the couplings between the
charmonium state and the hadronic molecular states are determined so as
to reproduce the observed mass of $X(3872)$ and the attraction between $D$
and $\bar D^{\ast}$ is determined so as to be consistent with the
observed $Z_b^{\pm ,0}$(10610) and $Z_b^{\pm ,0}$(10650) masses. 
Isospin symmetry breaking is introduced through the mass 
difference between the neutral and charged $D$ mesons. 

We have obtained the structure of $X(3872)$, which consists of 
6\% $\chi_{c1}(2P)$ charmonium, 69\% isoscalar $D \bar D^{\ast}$ molecule and 26\% isovector $D \bar D^{\ast}$ molecule.
This explains many of the observed properties of $X(3872)$, such as
isospin symmetry breaking, the production rate in the $p \bar p$
collision, the non-existence of the
 $\chi_{c1}(2P)$ peak predicted by
the quark model, and the absence of charged $X$.

We have further introduced the $J/\psi \rho$ and $J/\psi \omega$ channels
into the structure of the $X(3872)$ in Ref.~\cite{Takeuchi:2014rsa}.
The energy-dependent decay widths of the $\rho$ and $\omega$ mesons
have been introduced. The spectrum has been calculated up to 4 GeV. 
We have obtained very narrow $J/\psi \rho$ and $J/\psi \omega$ peaks
at around the $D^0 \bar D^{\ast 0}$ threshold, which is consistent with
the observation.

The $I(J^P) = 1(1^-), 1(3^-)$, and $1(5^-)$ $uudc \bar c$
pentaquarks have been investigated by the quark cluster model in
Ref.~\cite{Takeuchi:2016ejt}. This model, which reproduces the mass
spectra of the color-singlet $S$-wave $q^3$ baryons and $q \bar q$ mesons,
also enables us to evaluate the quark interaction in the color-octet
$uud$ configurations. It has been shown
that the color-octet isospin-$\frac{1}{2}$ spin-$\frac{3}{2}$ $uud$
configuration gains an attraction. The $uudc \bar c$ states with this 
configuration have structures around the $\Sigma_c^{(\ast)} \bar
D^{(\ast)}$ thresholds: one bound state, two resonances, and one 
large cusp are found. We have argued that the negative parity pentaquark 
found by the LHCb experiments may be given by these structures.


} 

%% file: zou.tex
\vfill\eject
\section{Early papers on \boldmath $P_c$ and on exotic nature of $N^*(1535)$}
\centerline{\large\bf Bing-Song Zou}
\centerline{CAS Key Laboratory of Theoretical Physics, Institute of Theoretical Physics, Chinese Academy of Sciences}
\centerline{Zhong Guan Cun East Street 55, Beijing 100190, China}
\centerline{School of Physical Sciences, University of Chinese Academy of Sciences, Beijing 100049, China}
\centerline{School of Physics, Central South University, Changsha 410083, China}
\centerline{\tt zoubs@itp.ac.cn}
\hfill\break

\begin{itemize}

\item First papers~\cite{Wu:2010jy,Wu:2012md} predicting the hidden charm $P_c$ and $P_{cs}$
penta-quark states above 4 GeV as hadronic molecules and suggesting to
look for them through their decays to $p J/\psi$ and $\Lambda J/\psi$,
respectively. The predicted states were observed by LHCb experiment a
few years later.

\item Paper demonstrating from BES data on $J/\psi\to\bar pK^+\Lambda$ and COSY data on $pp\to pK^+\Lambda$ that the $N^*(1535)$ has a large coupling to $K\Lambda$ and hence a large mixture of $\bar ssuud$ components~\cite{Liu:2005pm}. Extending from the hidden strangeness to hidden charm naturally leads to the expectation for the existence of $P_c$ states~\cite{Wu:2010jy,Wu:2012md}.

\end{itemize}

%% file: chen.tex
\vfill\eject

\section{QCD sum rules studies of exotic tetraquark states}
\centerline{\bf\large Hua-Xing Chen}
\centerline{School of Physics, Southeast University, Nanjing 210094, China}
\centerline{\tt hxchen@seu.edu.cn}
\hfill\break
%
%
In QCD sum rule study we need to construct the interpolating current $J$, which couples to the physics state $X$ we want to investigate. However, their relation is still not fully understood: a) the current $J$ sees only the quantum number of $X$, so it may also couple to some other physical states having the same quantum number; b) we can sometimes construct more than one currents, all of which couple to the same state $X$.

In Ref.~\cite{Chen:2006hy} we systematically constructed all the $u d \bar s \bar s$ interpolating currents of $J^{PC} = 0^{++}$, and used them to perform QCD sum rule analyses. We found five local currents in the diquark-antidiquark form ($[qq][\bar q \bar q]$) and ten local currents in the meson-meson form ($[\bar q q][\bar q q]$). We related them through the Fierz transformation, and verified that there are five independent ones.

An easier subject was studied in Ref.~\cite{Chen:2018kuu}, where we found only two independent $ss\bar s \bar s$ interpolating currents of $J^{PC} = 1^{--}$:
\begin{eqnarray}
&& \eta_{1\mu} = s_a^T C \gamma_5 s_b \bar{s}_a \gamma_\mu \gamma_5 C \bar{s}_b^T - s_a^T C \gamma_\mu \gamma_5 s_b \bar{s}_a \gamma_5 C \bar{s}_b^T,
\label{def:eta1}
\\ && \eta_{2\mu} = s_a^T C \gamma^\nu s_b \bar{s}_a \sigma_{\mu\nu} C \bar{s}_b^T - s_a^T C \sigma_{\mu\nu} s_b \bar{s}_a \gamma^\nu C \bar{s}_b^T.
\label{def:eta2}
\end{eqnarray}
Here $a$ and $b$ are color indices, $C = i\gamma_2 \gamma_0$ is the charge-conjugation matrix, and the superscript $T$ represents the transpose of Dirac indices only. We also constructed four meson-meson $(\bar ss)
(\bar s s)$ currents of $J^{PC} = 1^{--}$, which can be related to the above $\eta_{1\mu}$ and $\eta_{2\mu}$ through the Fierz transformation.

We calculated their diagonal terms
\begin{eqnarray}
\langle 0 | T \eta_{1\mu}(x) { \eta_{1\nu}^\dagger } (0) | 0 \rangle~~~{\rm and }~~~\langle 0 | T \eta_{2\mu}(x) { \eta_{2\nu}^\dagger } (0) | 0 \rangle \, ,
\label{eq:diagona}
\end{eqnarray}
as well as their off-diagonal term
\begin{eqnarray}
\langle 0 | T \eta_{1\mu}(x) { \eta_{2\nu}^\dagger } (0) | 0 \rangle \, .
\label{eq:offdiagona}
\end{eqnarray}
Based on the obtained results, we further constructed two (almost) non-corrected currents $J_{1\mu}$ and $J_{2\mu}$. We assumed that they couple to two different states, and calculated their masses to be
\begin{eqnarray}
M_{J_1} &=& 2.41 \pm 0.25  {\rm~GeV} \, ,
\\ M_{J_2} &=& 2.34 \pm 0.17  {\rm~GeV} \, .
\end{eqnarray}
The latter one suggests that $J_{2\mu}$ may couple to the $Y(2175)$, while the former non-corrected one suggests that the $Y(2175)$ may have a partner state with the mass $\Delta M = 71^{+172}_{-~48}$ MeV larger.

A recent BESIII experiment~\cite{BESIII:2021lho} studied the process $e^+ e^- \rightarrow \phi \pi^+ \pi^-$ and confirmed that the $Y(2175)$ has a partner state, labelled $X(2400)$, with the mass $M = 2298^{+60}_{-44} \pm 6$~MeV and width $\Gamma = 219^{+117}_{-112} \pm 6$~MeV.

%% file: meissner.tex
\vfill\eject
\section{Key papers on exotic hadrons by Ulf-G.~Mei\ss{}ner}

\centerline{Helmholtz-Institut für Strahlen- und Kernphysik and Bethe Center for Theoretical Physics,}
\centerline{Universität Bonn, D-53115 Bonn, Germany.}
\centerline{Institute for Advanced Simulation, Institut für Kernphysik and Jülich Center for Hadron Physics,}
\centerline{Forschungszentrum Jülich, D-52425 Jülich, Germany.}
\centerline{\tt meissner@hiskp.uni-bonn.de}
\hfill\break
\begin{itemize}
\item
Many exotic states close to two-particle threshold qualify as
hadronic molecules, that is loosely bound multi-quarks states
of two meson, a meson and a baryon or two baryons. In the
review \cite{Guo:2017jvc} we discuss the theory underlying such type of hadrons
and possible experimental tests to scrutinize their nature.

\item
In Ref.~\cite{Meissner:2020khl}
 we discuss two-pole structures in QCD.
 The two-pole structure refers to the fact that particular single states in the spectrum as listed in the PDG tables are often two states. The story began with the $\Lambda(1405)$ and has recently been extended to meson resonances.
This appears to be a general phenomenon in coupled channel hadron-hadron
scattering, when at least two channels in the group-theoretical basis
are attractive. 

\item
The LHCb pentaquark  $P_c(4457)$  is consistent with earlier predictions of a 
$\Sigma_c \bar D^*$ molecule with $I = 1/2$. In Ref.~\cite{Guo:2019fdo} we point out that if
such a picture were true, one would have
$B(P_c(4457) \to J/\psi \Delta)/B(P_c(4457)\to J/\psi p)$
at the level ranging from a few percent to about 30\%. 
Such a large isospin breaking decay ratio is two to three orders of magnitude larger than that for normal hadron resonances. It is a unique feature of
the $\Sigma_c\bar D^*$ molecular model, and can be checked by LHCb. 

\end{itemize}

%% file: guo.tex
\vfill\eject

\color{black}
\section{Key papers by Feng-Kun Guo}
\centerline{CAS Key Laboratory of Theoretical Physics, Institute of Theoretical Physics}
\centerline{Chinese Academy of Sciences, Beijing 100190, China}
\centerline{School of Physical Sciences, University of Chinese Academy of Sciences, Beijing 100049, China}
\centerline{\tt fkguo@itp.ac.cn}
\hfill\break

\begin{itemize}
 \item Near-threshold states and kinematical singularities

  In Ref.~\cite{Dong:2020hxe}, we show that near-threshold structures show up as long as the $S$-wave interaction is attractive, and the structure is more pronounced for heavier hadrons and for stronger attraction.

  Triangle singularities can produce peaks mimicking resonances and may also enhance the production of near-threshold states. Their possible realizations in hadronic reactions and effects of threshold cusps are extensively reviewed in Ref.~\cite{Guo:2019twa}.

  I proposed to measure the binding energy of the $X(3872)$ making use of triangle singularity in Ref.~\cite{Guo:2019qcn}, and a very high precision can be reached with such an approach. The approach is general such that it may also be used to precisely measure the binding energy of other near-threshold particles.
\end{itemize}

%% file: hanhart.tex
\vfill\eject
{\color{black}
\section{Work on Exotic Hadrons with involvement of the J\"ulich group}
\centerline{\large \bf Christoph Hanhart}
\centerline{Forschungszentrum J\"ulich, Institute for Advanced Simulation,} 
\centerline{Institut f\"ur Kernphysik, and J\"ulich Center for Hadron
Physics, 52425 J\"ulich, Germany}
\centerline{\tt c.hanhart@fz-juelich.de}
\hfill\break

The focus of the research by the J\"ulich group, often in collaboration with the Bochum group and  a group from ITP, Beijing, 
on multiquark states is
  on hadronic molecules and their possible imprints
in experimental observables. A  good description of the activities of the J\"ulich group in this field
and  also a list of relevant references can be found in two review articles already  cited by other contributors to this white paper \cite{Guo:2017jvc,Brambilla:2019esw}. 

The problem is approached from three sides:
First of all, experimental signals for certain states are investigated using the famous Weinberg criterion,
generalised to unstable and virtual states. When applied to the $T_{cc}$ and to the $\chi_{c1}(3872)$, also known as
$X(3872)$, this criterion reveals for both states a very
strong evidence for a molecular nature~\cite{Baru:2021ldu}. 
 
Second, effective field theories are constructed in analogy to what is done in the application of 
chiral effective field theories to atomic nuclei.
In this kind of approach at leading order there appear counterterms, whose number is constrained mostly
by heavy-quark spin symmetry, as well as by the one-pion exchange that induces $S$-$D$ transitions.
The latter
 turn out to be very strong, especially in  a coupled-channel setting for doubly heavy hadronic molecules
 formed from $D^{(*)}\bar D^{(*)}$ or $B^{(*)}\bar B^{(*)}$ interactions, since the channel couplings
 naturally introduce typical momenta of the order of $\sqrt{2\mu\Delta M}\approx 500$ MeV, where $\mu$
 denotes the reduced mass of the two-meson system and $\Delta M$ the heavy quark spin symmetry violating
 $D-D^*$ and  $B-B^*$  
 mass difference, respectively.
The effect is reduced significantly for the $T_{cc}$, where the mass difference between the relevant channels
is driven by isospin violation. The latter kind of calculation is reported in 
in Ref.~\cite{Du:2021zzh} for the $T_{cc}$.
A systematic study of the regulator dependence of the results, which can be regarded as a numerical
implementation of the renormalization group equations, showed that in the general case an $S$-$D$ counterterm
must be promoted to leading order in order to arrive at a consistent effective field theory.
At next-to-leading order additional  energy-dependent counterterms, as well as two-pion
exchange contributions enter. 

Last but not least,
of particular interest for unraveling the nature of heavy states  are systematic studies of the
extent of various symmetries' violation.  Thus e.g. in Ref.~\cite{Cleven:2015era} it is demonstrated   
that
spin
symmetry violation is quite sensitive to the composition of the states. 
For example, in the molecular picture the lightest $J^{PC}=0^{-+}$ is predicted  about 100 MeV
above the mass of the $\psi(4230)$, also known as $Y(4230)$, where the latter appears
as $D_1 \bar D$ bound system, since $J=0$ in this scenario can only be reached in the
$S$-wave from $D_1\bar D^*$. In contrast, the lightest exotic hadrocharmonium state
is predicted to reside 100 MeV below the mass of the $Y(4230)$ --- a prediction emerging
from a certain mixing scenario necessary to explain the appearance of this vector
state in both  $h_c \pi \pi$ and in $J/\psi \pi \pi$ final states. Finally, in the compact tetraquark
scenario the lightest $0^{-+}$ multiquark is predicted in between.
More examples of this kind can be derived.
In addition the the extent of spin-symmetry violation, the extent of $SU(3)$-flavor violation
can also be an effective tool for studying the structure of exotic mesons, simply because hadronic molecules
(if they exist)
are located close to the production thresholds of their constituents and thus the extent of 
$SU(3)_f$ breaking within their multiplets is dictated by the amount of $SU(3)_f$ violation in the constituent $\bar Qq$ multiplets
forming the molecule. Such a connection is absent in the other scenarios.



%% file: swanson.tex
\vfill\eject

\section{Multicomponent Hadrons}
\centerline{\bf\large Eric S. Swanson}
\centerline{Department of Physics and Astronomy, University of Pittsburgh, Pittsburgh, PA 15260, USA}
\centerline{\tt swansone@pitt.edu}
\hfill\break
The existence and properties of multiquark---and multi-component---hadrons has been of interest since the beginnings of the quark model (and hence predates QCD). 
The situation is much more robust now  and moreover  the application of lattice field theory and effective field theory to the multiquark problem has matured and greatly assists  in developing understanding. In spite of these gains, much remains unknown. 

The $X(3872)$ is widely regarded as the first heavy multiquark state, with early work establishing its quantum numbers and decay modes before they were measured\cite{Swanson:2003tb}. This state triggered  much interest in heavy multiquark hadrons that  led to many advances in theoretical methods  and  also to the discoveries  of other heavy quark exotic states .  Amongst these are a collection of pentaquark states, that have been  interpreted as weakly bound $\Sigma_c^{(*)}\bar{D}^{(*)}$ states in \cite{Burns:2021jlu} . Another possibility is that some of these states are associated with dynamically generated singularities that are not due to resonances, such as threshold cusps or triangle diagram singularities\cite{Swanson:2014tra}. The latter is  a possibility that can also occur in hadronic reactions. 
Indeed, it is evident that the field is well beyond its early phase of hunting for isolated bumps and interpreting them in terms of Breit-Wigner amplitudes. Rather, reactions can involve many dynamical effects and can yield complex signals that require sophisticated analysis techniques. We are only at the beginning of developing the necessary tools for the explosion of information that we have now and the next years promise to reveal much about the properties of the strongly interacting part of the Standard Model.

%% file: pilloni.tex
\vfill\eject

\section{Line shape analyses and the nature of exotic states}
\centerline{\bf\large Alessandro Pilloni$^{1*}$ and Adam P. Szczepaniak,$^{2\dagger}$}
\centerline{\vrule width 0pt height 3ex
$^1$Università di Messina, I-98122 Messina and INFN Catania, I-95123 Catania, Italy}
\centerline{$^2$Enrico Fermi Institute and Department of Physics and}
\centerline{Center for Exploration of Energy and Matter, Indiana University, Bloomington, IN 47403, USA and}

\centerline{Department of Physics, Indiana University, Bloomington, IN 47405, USA and}

\centerline{Theory Center, Thomas Jefferson National Accelerator Facility, Newport News, VA 23606, USA}
\centerline{\tt $^*$†
alessandro.pilloni@unime.it \quad $^\dagger$§
aszczepa@indiana.edu}
\hfill\break
\subsection*{Line shape study of $e^+e^- \to J/\psi,\pi\pi$ and $\to D\bar D^* \pi$ data from BESIII and the nature of the  $Z_c(3900)$.}
The $Z_c(3900)$ peaks in $J/\psi\,\pi$ and enhances the $D\bar D^*$ cross section at threshold. 
Several interpretations have been put forward: it might be a bound or virtual state of $D \bar D^*$, that acquires a width  due to the coupling to $J/\psi\,\pi$; it might be a genuine QCD resonance; it might be a mere threshold cusp enhanced by the presence of a triangle singularity closeby; or a combination of all these.  
 The best candidate to produce a triangle cusp is the $D_1(2420)$  resonance in $D^* \pi$. 
Each of these interpretation is reflected in the analytic properties of the amplitude, 
which in turn affects the for studying its nature.

 We perform a coupled-channels study of the $e^+e^- \to J/\psi,\pi\pi$ and $\to D\bar D^* \pi$ data from BESIII 
 \cite{Pilloni:2016obd}. We write a unitarized model that takes into account possible rescattering with the
 bachelor particle in both channels. 
 We consider several amplitude parametrizations that favor different physical interpretations. 
 All the models fit the data reasonably well, with $\chi^2/\text{d.o.f.}$ ranging from $1.2$ to $1.3$. 
 A likelihood ratio test does not give rejections larger than $3\sigma$. 
 We conclude that present statistics prevents us from drawing any strong conclusions.

\subsection*{Line shape studies of $\Lambda_b^0\to J/\psi K^- p$ data from LHCb and interpretation of the $P_c(4312)$ as a virtual state. }
The discovery of two pentaquark resonances,
$P_c(4380)$ and $P_c(4450)$ in the $\Lambda_b^0\to J/\psi K^- p$
decay by LHCb in 2015 has stimulated the fantasy of theorists trying to pin down their nature. 
Later, with ten times more events, the $P_c(4450)$ signal
has been resolved into two peaks, $P_c(4440)$ and $P_c(4457)$,
and a new $P_c(4312)$ was discovered.
The latter is particularly interesting, as it is a very clean isolated structure,
peaking $5$~MeV below the $\Sigma_c^+\bar{D}^0$ threshold.
This calls for a natural explanation as a hadron molecule composed 
of the two particles. 
However, the interaction between the $\Sigma_c^+$ and the $\bar{D}^0$ can also generate a virtual state,
if the attraction is not strong enough to provide binding, as happens in di-neutron scattering.
The narrow ($\sim 10$~MeV) peak appears on top of
a smooth background, permitting a simplified analysis of the one-dimensional $J/\psi\,p$ 
invariant mass distribution.

We consider a two-channel model, $\Lambda^0_b \to K^- (J/\psi\,p)$ and $\to K^- (\Sigma_c^+\bar{D}^0)$. 
Since we focus on events around the $P_c(4312)$ peak only, far away from the $J/\psi\,p$ threshold,
the latter absorbs all the channels lighter than $\Sigma_c^+\bar{D}^0$. Similarly, the contributions 
from heavier channels can be absorbed by the real parameters of the scattering amplitude. 
At threshold one can use the scattering length approximation to parametrize the amplitude. 
The sign of one of the amplitude parameters indicates whether the state is virtual or bound. 

The data favors a virtual state interpretation, with
$M_P = 4319.7 \pm 1.6$~MeV and $\Gamma_P = -0.8 \pm 2.4$~MeV, 
where a negative width is conventional for virtual states. 
Consistent results are obtained with the three LHCb datasets \cite{Fernandez-Ramirez:2019koa}.
The same analysis was repeated using a deep neural network, trained
on four classes of lineshapes, representative of state nature (bound or virtual) 
and the Riemann sheet on which the pole is located.
Again, the analysis heavily favors a virtual state interpretation \cite{Ng:2021ibr}, 
validating use of Machine Learning in hadron spectroscopy.

%% file: prelovsek.tex
\vfill\eject

\section{Exotic Hadrons in Dynamical Lattice QCD}
\centerline{\bf\large Sasa Prelovsek}
\centerline{Faculty of Mathematics and Physics, University of Ljubljana}
\centerline{and Jozef Stefan Institute, 1000 Ljubljana, Slovenia}
\centerline{\tt sasa.prelovsek@ijs.si}
\hfill\break

All works below refer to dynamical  lattice QCD simulations of exotic
hadrons, authored by Sasa Prelosek and collaborators.

\begin{itemize}
\item [1.] Ref. \cite{Prelovsek:2013cra} was the first lattice study that finds an evidence for $X(3872)$. It extracted the $D\bar D^*$ scattering amplitude near threshold and found a bound state pole in it. The pole appears just slightly  below threshold and is related to $X(3827)$ with $J^P=1^+$ and $I=0$.

\item [2] In Ref. \cite{Prelovsek:2020eiw} the  coupled  $D\bar D-D_s\bar D_s$ scattering renders the expected conventional charmonia  with $J^{PC}=0^{++},~2^{++}$ and in addition two unconventional scalar states just below $D\bar D$ and $D_s\bar D_s$ thresholds; the first has not been discovered experimentally (however the reanalysis of exp data supports it), while the  second one might be related to $X(3915)/\chi_{c0}(3930)$  

%
\item[3.] In Ref.~\cite{Padmanath:2022cvl} 
the doubly charm tetraquark with flavor $cc\bar u\bar d$ and isospin
$I\!=\!0$ is investigated by calculating the $DD^*$ scattering
amplitude with lattice QCD.
A virtual bound state pole in the $DD^*$ scattering amplitude with $l=0$
is found $9.9_{-7.1}^{+3.6}~$MeV below $DD^*$ threshold. This pole is likely related to
the doubly charmed tetraquark discovered by LHCb  \cite{LHCb:2021vvq,LHCb:2021auc} less than $1~$MeV below
$D^0D^{*+}$ threshold.

\end{itemize}

%% file: mathur.tex
\vfill\eject
\section{$\bar Q \bar Q q_1 q_2$ tetraquarks on the lattice}
\centerline{\bf\large Nilmani Mathur}
\centerline{Department of Theoretical Physics, Tata Institute of Fundamental Physics}
\centerline{Homi Bhabha Road, Mumbai 400005, India}
\centerline{\tt nilmani@theory.tifr.res.in}

\hfill\break

\newcommand\bef{\begin{figure}}
Being a first principles method with quantifiable and improvable
systematic errors, lattice QCD is a natural choice for studying the
structures and interactions of subatomic particles.  Recent
advancements in computing the excited state energy spectra in the
finite-volume as well as in extracting the hadron-hadron scattering
amplitudes, have made lattice QCD an appealing tool to investigate the
exotic particles. Lattice QCD practitioners are already performing
such studies, and in fact, initial results from several groups have
suggested the existence of strongly bound spin-1 doubly bottom
four-quark states with the valence quark contents of
$\bar{b}\bar{b}ud$ and $\bar{b}\bar{b}us$.

In Ref. \cite{Junnarkar:2018twb} we carried out a detailed calculation
on four-quark states with the valence quark contents
$\bar{Q}\bar{Q}q_1q_2 ; \,\,Q \in b,c$ and $q_1, q_2 \in u(d),s$.  We
varied the non-heavy quark masses ($m_q$) over a wide range and
extracted the corresponding finite-volume energy spectra of various
four-quark flavor combinations. The ground state energies with respect
to the elastic threshold ($\Delta E$) are shown in Fig
\ref{fig:summary_spinone} for a number of doubly heavy four-quark
configurations and compared those with other available results. These
results suggest the presence of an energy level of about 100-150 MeV
below the elastic threshold ($BB^*$) for $\bar{b}\bar{b}ud$, which can
be associated with its binding energy assuming the finite-volume effects
are small in a system of two heavy mesons.  For $\bar{b}\bar{b}us$,
the respective threshold and the predicted binding energy are $BB_s^*$
and 70-100 MeV. We also find an interesting QCD-dynamics that heavier
the heavy quark masses and lighter the light quark masses the stronger
is the binding for a doubly heavy four-quark system. For their charm
siblings, $\bar{c}\bar{c}ud$, $\bar{c}\bar{c}us$, we found an energy
level of about $23 \pm11 $ MeV and $8\pm 8$ MeV below their respective
elastic thresholds, $DD^*$ and $DD_s^*$. Although, it is tempting to
relate this below-the-threshold energy level with the recently
discovered $T_{cc}^{+}$, a rigorous inference on signatures for a
$T_{cc}^{+}$ state from lattice calculations requires extraction of
the $DD^*$ scattering amplitudes from the finite volume spectrum,
followed by a pole search in the complex energy plane. However, such
calculations involve detail analysis and are much more
computation intensive. We plan to perform such studies in the near future.

Recently we also studied $\bar{b}\bar{c}ud$ and $\bar{b}\bar{c}us$,
and preliminary results indicate the presence of an energy level
approximately $20-40$ MeV below the respective elastic thresholds ,
$B^*D$ and $B^*_sD$ \cite{Padmanath:2021qje}.  We believe the absence
of such energy levels in other lattice calculations is related to the
cut-off effects from the heavy quarks in relatively coarse lattice
spacings. A detailed calculation involving finite-volume study is
underway for these systems for which experimental search may also be
possible in the near future.

For the spin-0 doubly heavy four-quark systems we do not find any
energy level below their respective elastic thresholds which suggest no
strong binding for such systems.

\begin{figure}[h]
\begin{center}
	\includegraphics[width=0.75\linewidth]{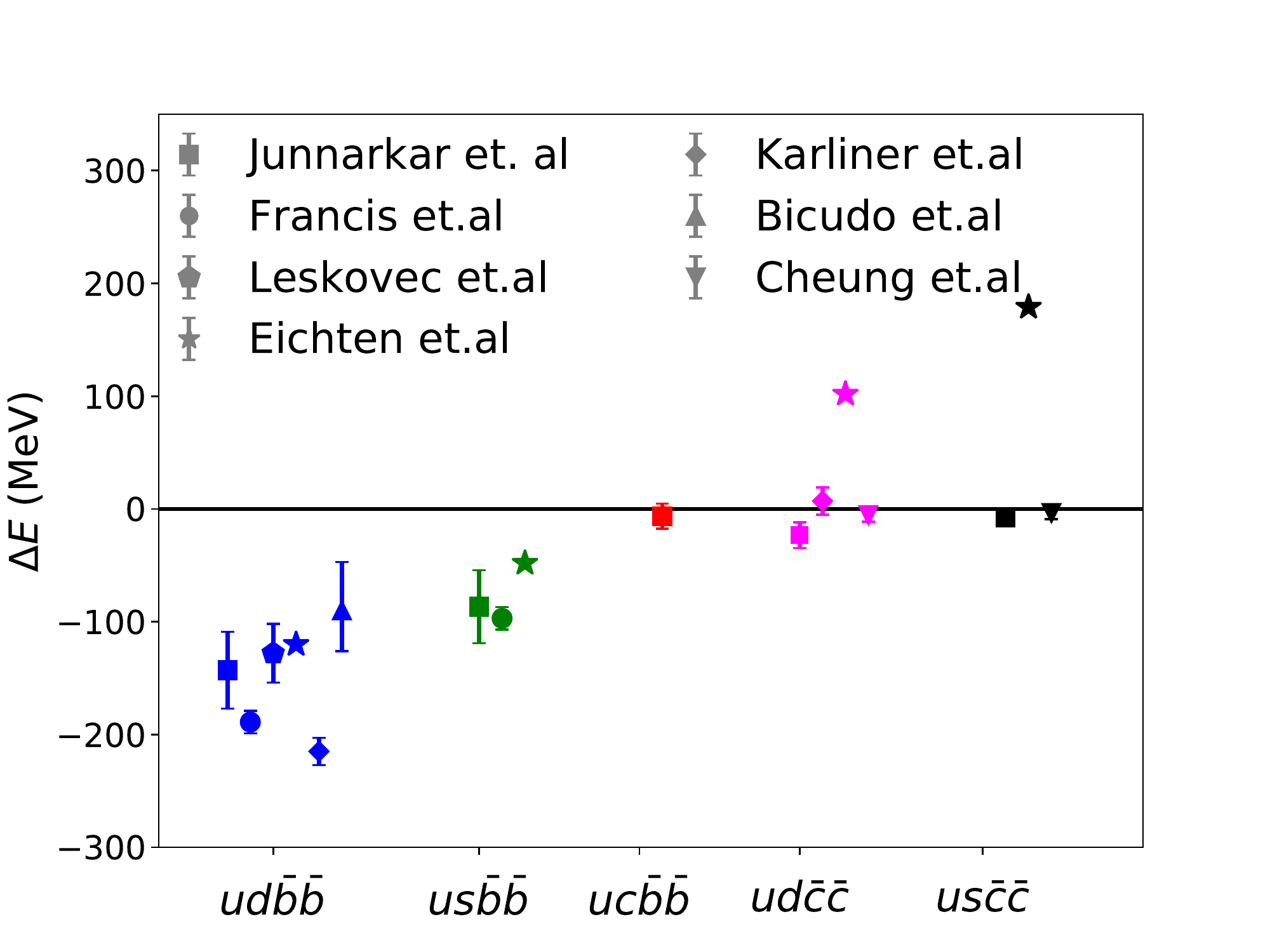}
\end{center}
\caption{\label{fig:summary_spinone}Comparison of global results on the spin-1 doubly bottom and charm four-quark
states with various flavor combinations. $\Delta E$ is the energy difference between the ground state and the 
lowest elastic threshold. Various flavor combinations represented on the horizontal axis are color coded as: 
blue, green, red, magenta and grey for the state
$ud\bar{b}\bar{b}, \ us\bar{b}\bar{b}, \ uc\bar{b}\bar{b}, \ ud\bar{c}\bar{c}$ and $us\bar{c}\bar{c}$, respectively.}
\end{figure}

Beside the four-quark states we also investigate six-quark bound
states. In Ref. \cite{Junnarkar:2019equ} heavy dibaryons were studied
and the results suggest that strong-interaction-stable deuteron-like
heavy dibaryons can exist if at least two of the quarks in a dibaryon
have heavy flavors. Unlike the four-quark systems it was found that
for stronger binding it is preferable to have all quark masses to be
heavier and naturally $\Omega_{bbb}\Omega_{bbb}$ has the strongest binding. A
recent lattice calculation has also reported a weakly bound state for
charmed dibaryons with six charm quarks. However, our findings on the
three-flavored $H$-like dibaryons suggest no strong binding for any
combinations of quark masses at their physical values including also the
heavier quark masses.




%% file: francis.tex
\vfill\eject
\section{Diquarks and doubly heavy tetraquarks from LQCD }

\centerline{\bf\large Anthony Francis$^{a,b,c,*}$ and Randy Lewis$^{d,\dagger}$} 
\vrule width 0pt height 3ex
\centerline{$^a$Albert Einstein Center, Universität Bern, Sidlerstrasse 5, 3012 Bern, Switzerland}
\centerline{$^b$Institute of Physics, National Yang Ming Chiao Tung University, 30010 Hsinchu, Taiwan}
\centerline{$^c$Theory Department, CERN, 1201 Geneva, Switzerland}
\centerline{$^d$Dpt of Physics and Astronomy, York University, Toronto, Ontario, M3J 1P3, Canada}
\centerline{\tt $^*$anthony.francis@cern.ch \quad $^\dagger$randy.lewis@yorku.ca}
\hfill\break
  \newcommand{\udbb}{$bb \bar{u}\bar{d}$ }
  \newcommand{\lsbb}{$bb \bar{\ell}\bar{s}$ }
  \newcommand{\udcb}{$bc \bar{u}\bar{d}$ }

\subsection*{Diquark properties from full QCD lattice simulations }

The concept of diquarks is almost as old as the quark model, and actually
pre-dates QCD \cite{Gell-Mann:1964ewy}, \cite{Ida:1966ev}. 
In spite of the long history of successful phenomenological models for
low-lying baryons and exotics, experimental evidence has been
difficult to obtain, however.

Simultaneously, even though diquarks are well founded in QCD,
non-perturbative, ab-initio results have been scarce, in particular from
lattice QCD. The reason is that diquarks are coloured objects, i.e. not
gauge-invariant, and the lattice cannot access them easily.

In \cite{Francis:2021vrr} we address this issue by forming a gauge-invariant probe to diquark properties through embedding them in hadrons that contain a single static (=infinitely heavy) quark. This quark can be cancelled exactly in mass differences. Furthermore this configuration can be used to define a measure for the diquark structure through density-density correlations.

Diquark-diquark and diquark-quark mass differences have been called
``fundamental characteristics of QCD"
\cite{Jaffe:2004ph} and are interesting in their own right. 
We perform lattice calculations of these differences at a single lattice
spacing at fixed volume with a range of light quark masses from 707
to 164~MeV and find very good agreement with updated, phenomenological
estimates.
Going further and having validated our approach by this success, in the main
body of our work we study the spatial correlations within a diquark. We
successfully establish the unique attractive interaction in the ``good"
diquark channel. Polarisation effects due to the presence of the static
quark are not observed and the good diquark wave function is seen to be
spherical.  Furthermore, the decay of the spatial correlation between the
$q$-$q$ pair with distance enables us to determine the size of the diquark. We
find it is $\sim 0.6$ fm in diameter which entails it is of hadronic size.
We attach representative figures of our key findings in Fig.~\ref{fig:dens_panel} and \ref{fig:rad_panel}.

Moreover,  in Ref.~\cite{Francis:2021vrr}  Fig.1 top
panel, shows the dependence on $m_\pi$ of the good
diquark $0^{+}$ mass versus the $1^{+}$ one, providing support to the
phenomenological diquark approach, 
which focuses on the good $ud$ $0^+$ diquark, since the  good $ud$ $0^{+}$
diquark  is significantly lighter, 100-200 MeV 
below the bad $ud$ $1^{+}$ diquark.

\begin{minipage}[t!]{0.45\textwidth}
\centering
{\includegraphics[width=0.99\columnwidth]{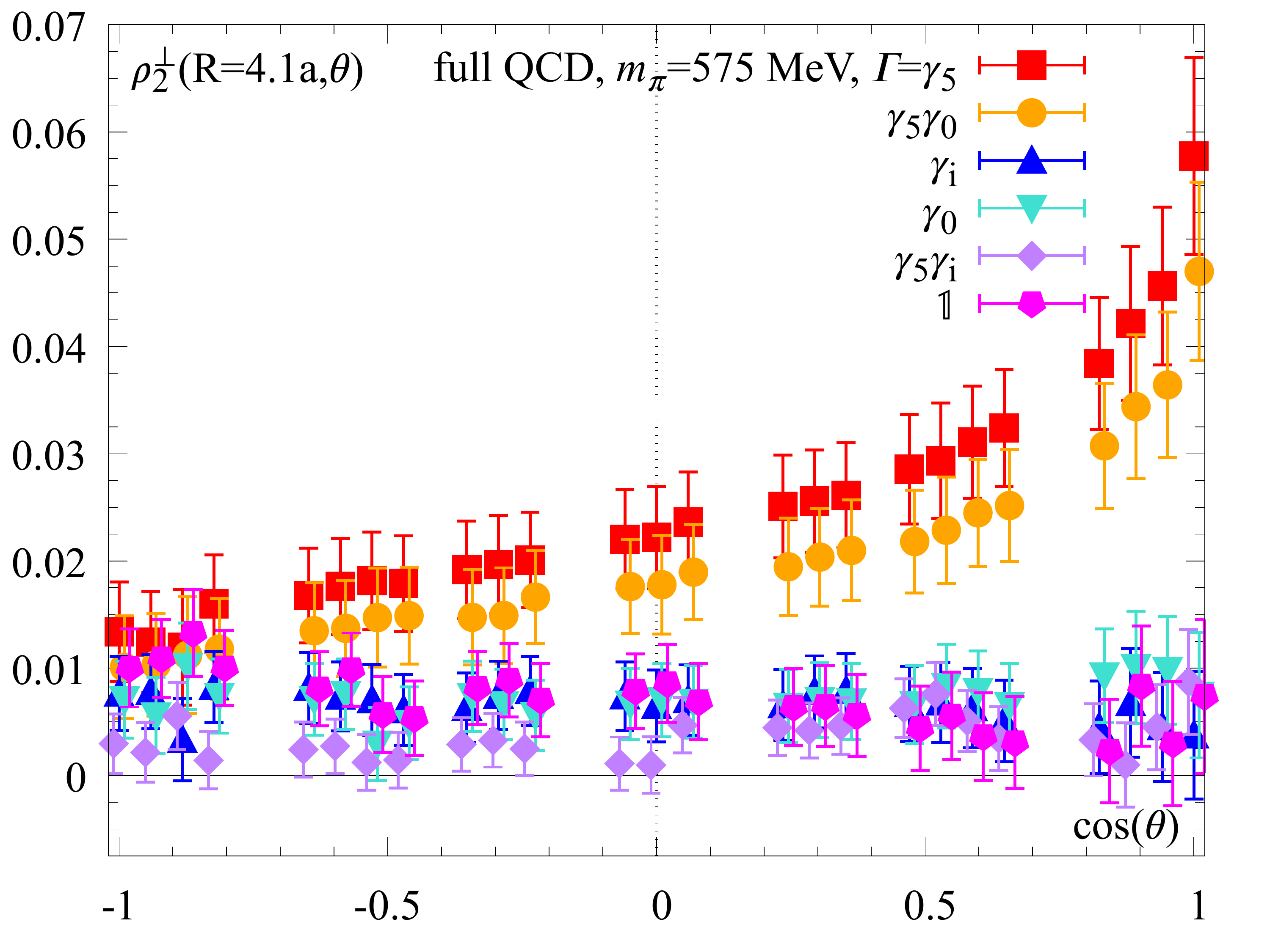}}
{\includegraphics[width=0.99\columnwidth]{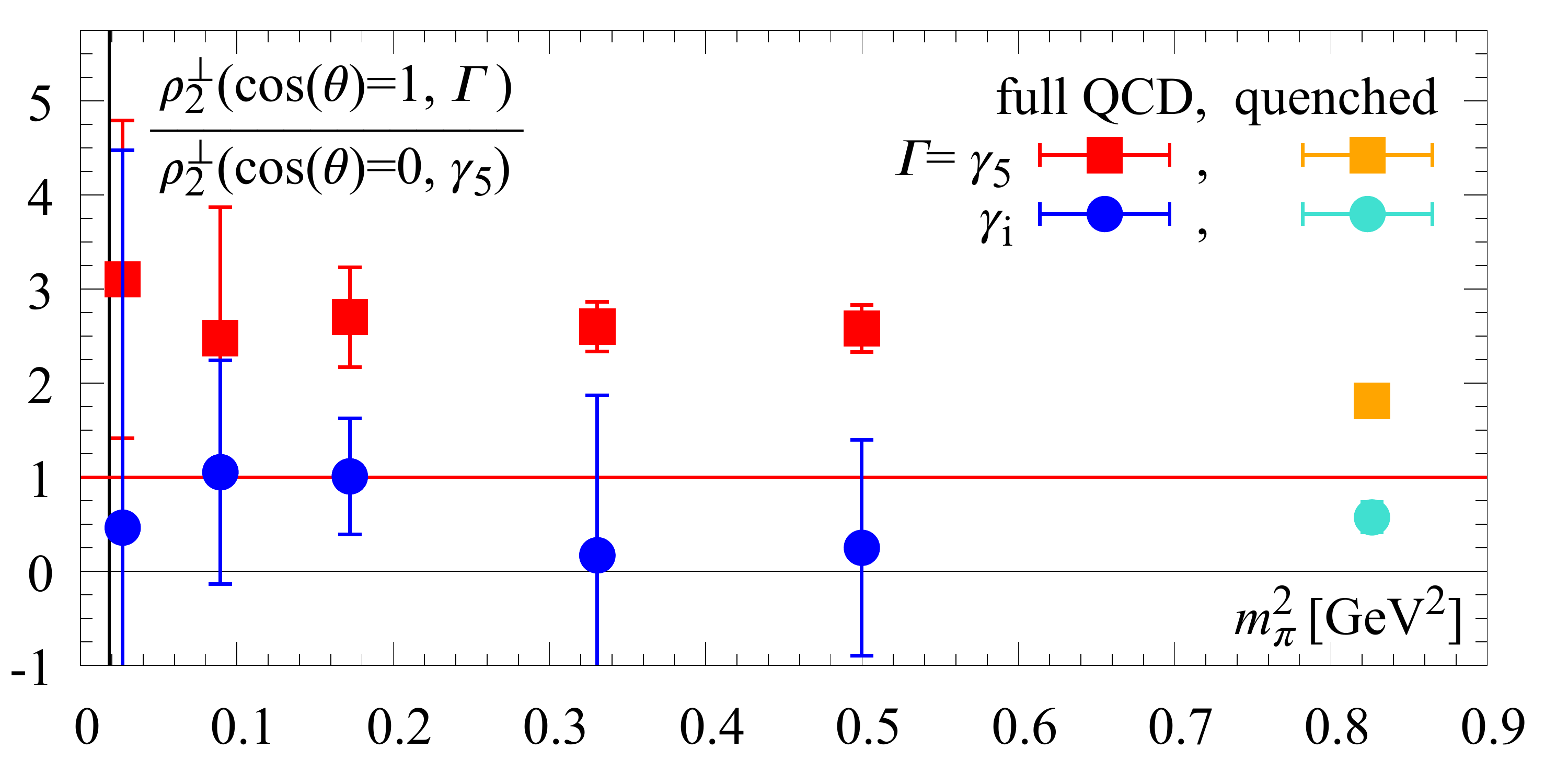}}

{\vspace{1ex}\includegraphics[width=0.8\columnwidth]{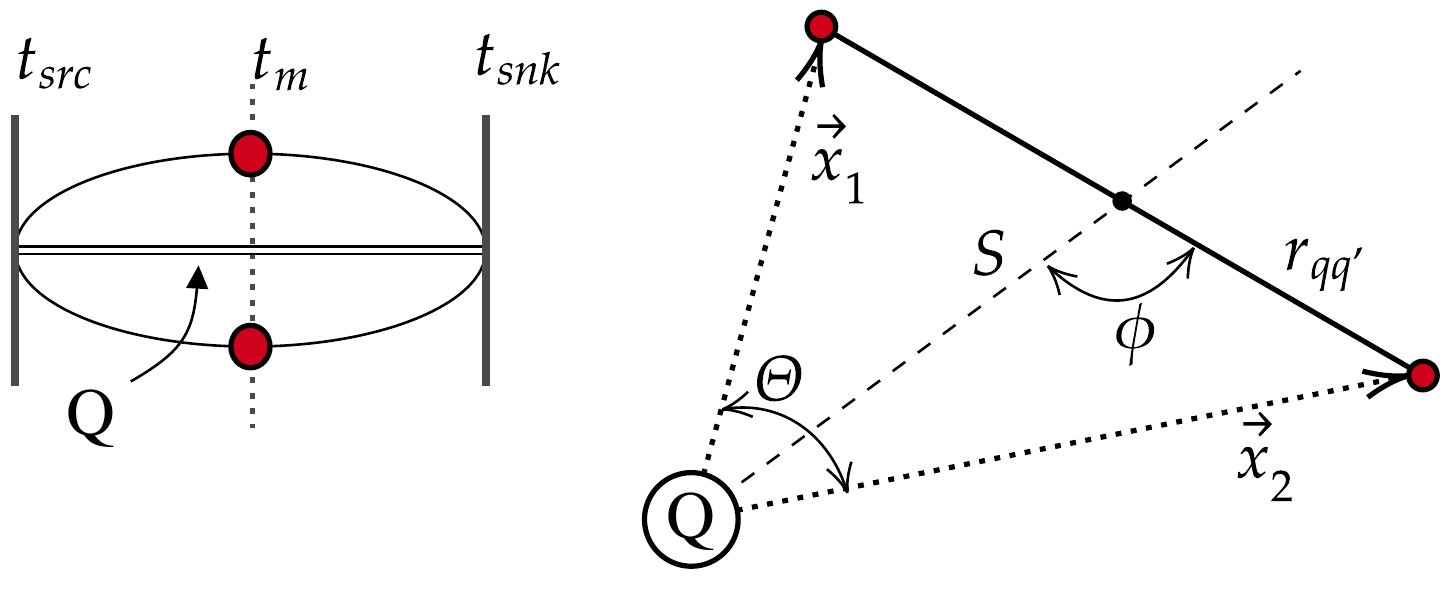}}
\captionof{figure}{{
{\it Diquark attractive effect. (Top) The density-density correlators
$\rho_2^\perp (R=4.1 a,\Theta ,\Gamma )$ versus 
$cos(\Theta )$ at $m_\pi=575~\rm{MeV}$. 
(Middle) The ratio
$\rho_2^\perp (R,\Theta =0,\Gamma )/
\rho_2^\perp(R,\Theta =\pi /2, \Gamma =\gamma_5)$
versus $m_\pi^2$. Values above/below 1 for 
the red/blue points signal
an attraction in the good diquark that is 
absent for the bad diquark. The vertical line
denotes physical $m_\pi$. 
(Bottom) Sketch of the 
density correlators: 2D temporal
view (left) and current insertions, spatial view (right).
} 
}}
\label{fig:dens_panel}
 \end{minipage}
\hspace{2ex}
 \begin{minipage}[t!]{0.45\textwidth}
\centering
{\includegraphics[width=0.99\columnwidth]{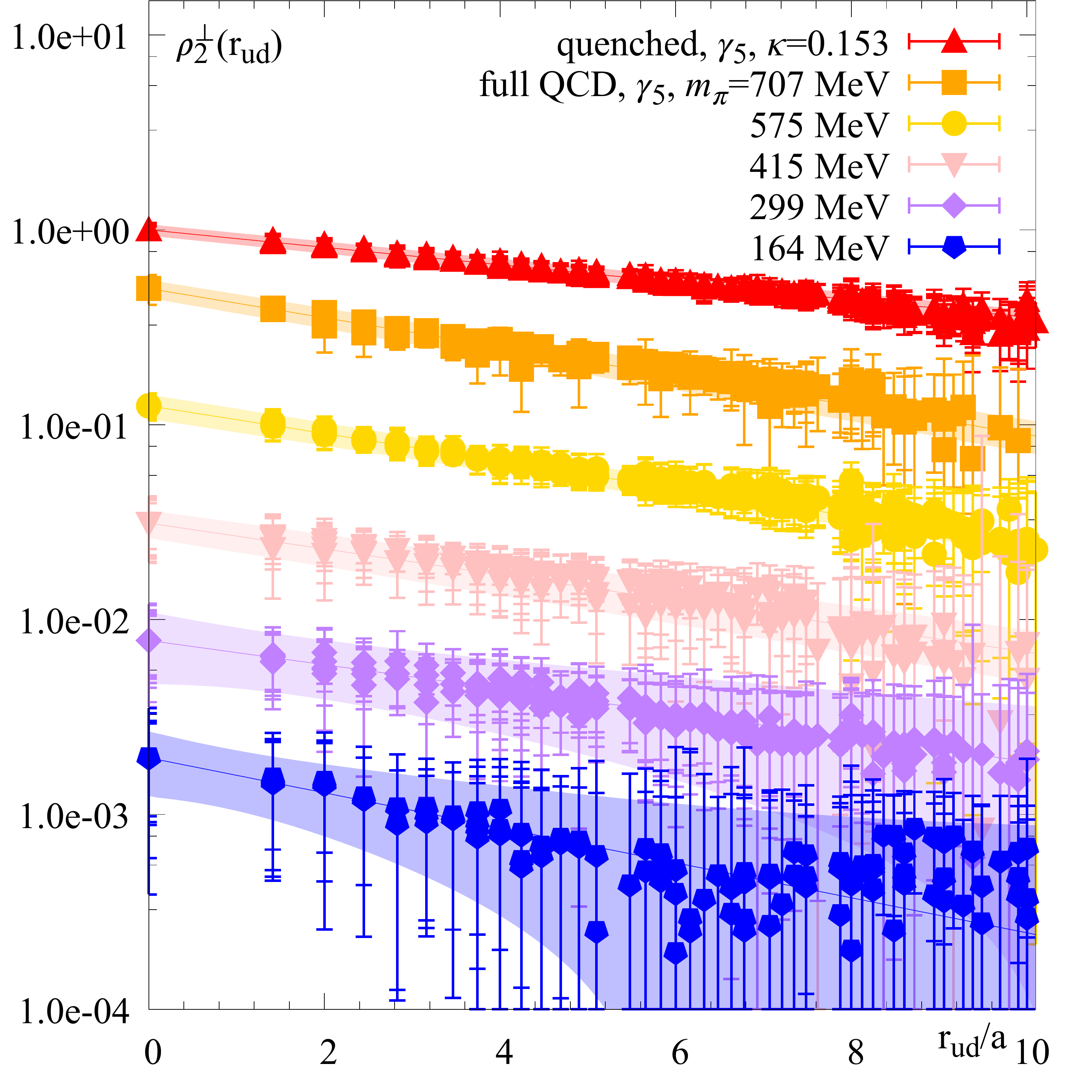}}
\vspace{-8pt}
{\includegraphics[width=0.99\columnwidth]{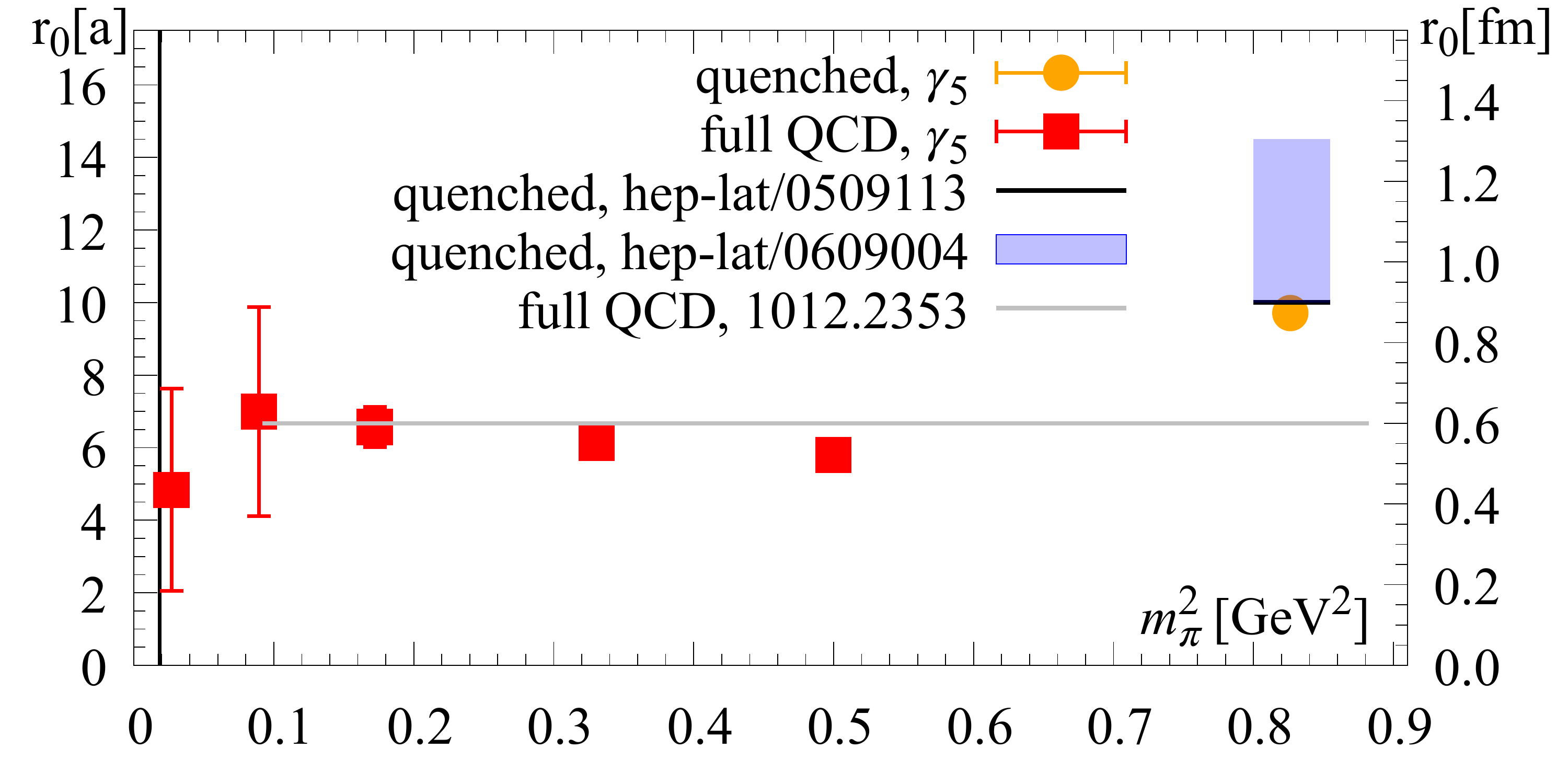}}
\captionof{figure}{{ 
\it
Good diquark size. (Top) 
Exponential decay with $r_{qq^\prime}$ 
of the 
$\rho_2^\perp (R,\Theta )$. 
Each $m_\pi$ has its own color.
Data sets have been normalised at 
$r_{qq^\prime}=0$ and offset vertically. 
Results for all available $R$ are shown 
together in one coloured set. Each coloured band
comes from the combined fit used to determine the diquark size
$r_0(m_\pi^2)$. (Bottom) 
Resulting good diquark size $r_0$ versus
$m_\pi^2$, compared to results from the 
literature. The vertical line denotes physical
$m_\pi$.  
}
\label{fig:rad_panel}
}

\end{minipage}

\vfill\eject
\subsection*{Doubly heavy tetraquarks in lattice QCD }

On the lattice there is a technical difficulty in studying exotic hadrons,
such as the $X(3872)$, because their signal is buried as excitation
in the short distance regime of a correlation function of the given
quantum numbers. Ground states dominate Euclidean correlation functions
at asymptotic times as the higher excitations are suppressed. This
suppression makes it difficult to access these higher states in the
spectrum.

Motivated by this outlook we decided to search for exotic
candidates that have the distinct feature of being ground states \cite{Francis:2016hui}.
Based on the concepts of heavy quark spin
symmetry and the good diquark attractive effect, we set our goal and
performed a lattice calculation of a doubly heavy tetraquark with quantum
numbers $J^P=1^+$ and flavor contents \udbb and \lsbb, ${\ell}=u,d$. We
found both
candidates exhibit significant binding below the relevant thresholds
and are strong-interaction stable.

In \cite{Francis:2018jyb} we extended this analysis by varying the heavy
quark mass components and could verify that the candidates' binding
energies indeed vary according to the phenomenological picture in mind,
giving some tentative, yet not conclusive, evidence of its validity. In
the same work we found indication that \udcb could also be a bound state,
but much shallower. The fate of this state is currently undetermined, as
our recent follow-up study did not observe signal for it any more
\cite{Hudspith:2020tdf}. Additionally, in this study we performed a
survey
of possible candidates with different flavor combinations in addition
to the quantum number channel $J^P=0^+$. We did not observe deep binding
for more candidates aside of the already established \udbb and \lsbb, in
$J^P=1^+$. The status of shallow bound states or even resonances is not
clear from this study and requires further work in the future.


\vspace{5ex}
\vfill\eject

%% file: ferretti.tex
\vfill\eject
{\color{black}
\section{Exotic hadron spectroscopy and decays}
\centerline{\bf\large Jacopo Ferretti$^{1*}$
and Elena Santopinto$^{2\dagger}$
} 

\centerline{\vrule width 0pt height 3ex
$^1$Physics Dpt., University of Jyv\"askyl\"a, P.O.B. 35 (YFL), 40014 Jyv\"askyl\"a, Finland}}
\centerline{$^2$INFN Sezione di Genova, I-16146 Genova, Italy}
\centerline{\tt $^*$jferrett@jyu.fi \quad $^\dagger$elena.santopinto@ge.infn.it}
\hfill\break

{\bf Hidden-charm tetra- and pentaquarks in the compact tetraquark and hadro-charmonium models.}
By making use of SU(3) flavor symmetry considerations, in Ref. \cite{Ferretti:2020ewe} it was argued that the existence of $Z_{\rm c}$ and $P_{\rm c}$ exotics necessarily implies the existence of $Z_{\rm cs}$ and $P_{\rm cs}$ hadrons.
The spectra of hidden-charm tetraquarks and pentaquarks with strangeness were also computed by means of the hadro-charmonium and relativized diquark models, \cite{Ferretti:2020ewe}.
The results presented here \cite{Ferretti:2020ewe} anticipated by almost a year the LHCb and BESIII findings of $Z_{\rm cs}$ states and, therefore, were cited in the LHCb and BESIII papers on the $Z_{\rm cs}(3985)$ and $Z_{\rm cs}(4003)$ tetraquarks.

In \cite{Santopinto:2016pkp}, which used only symmetry considerations and an equal- spaced mass formula, the hidden-charm pentaquark with strangeness,  $P_{cs}^0(4459)$,  was predicted 3 years in advance; this suggested  that this state should be sought in the $J/\Psi \Lambda$ channel \cite{Santopinto:2016pkp}, which is the same channel where the $P_{cs}^0(4459)$ was observed by LHCb this year \cite{LHCb:2020jpq}. In \cite{Santopinto:2016pkp}, it was argued that, if the observed exotics are 
molecules, one would not necessarily observe complete
$SU(3)_f$ multiplets.  In compact models, by contrast,
the emerging $SU(3)_f$ multiplets will be equally
spaced.
\\

{\bf Fully-heavy tetraquarks: spectroscopy and decays.}
In Ref. \cite{Anwar:2017toa}, the masses of the fully-heavy (four-$c$ and four-$b$) 
ground-state tetraquarks were calculated by means of a relativized diquark model. 
A non-relativistic Hamiltonian with Coulomb-like interaction and mass inequality relations were also derived.
\begin{figure}[htbp] 
\centering 
\includegraphics[width=5cm]{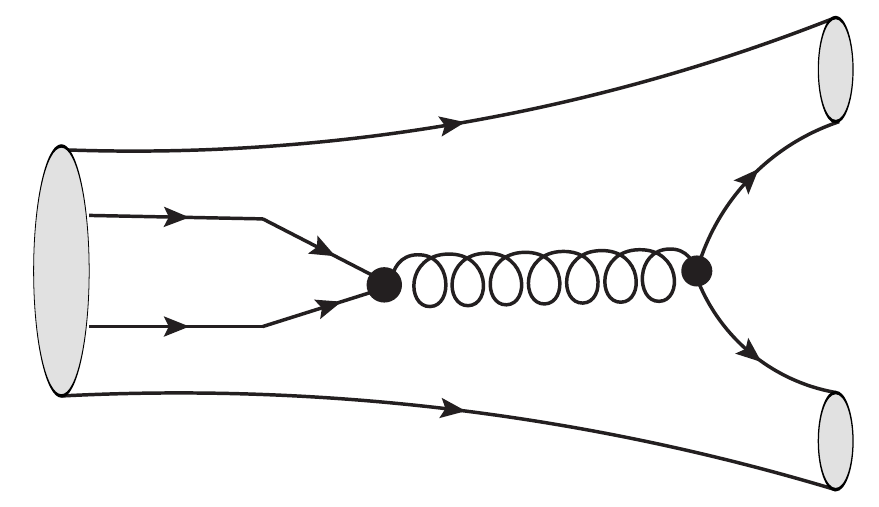}
\caption{Quark-level description of hadronic decays $X_{bb\bar{b}\bar{b}} \to M_1 \bar{M}_2$, where $M_1$ and $\bar{M}_2$ are the allowed spin-parity and phase-space bottom- and anti-bottom mesons, respectively. Picture from Ref. \cite{Anwar:2017toa}; Springer Nature copyright.} 
\label{fig:4Q-decays}
\end{figure}
Moreover, the total decay widths of ground-state four-$c$ and \hbox{four-$b$} tetraquarks were estimated on the basis of phenomenological considerations. As shown in the diagram in Fig. \ref{fig:4Q-decays}, the decays proceed via $Q\bar Q$ gluon-annihilation; the calculated decay widths are $\Gamma(X_{bb\bar{b}\bar{b}}) =\mathcal{O}(50~\textrm{MeV})$ and $\Gamma(X_{cc\bar c\bar c}) = \mathcal{O}(100~\textrm{MeV})$ \cite{Anwar:2017toa}.

In Ref. \cite{Bedolla:2019zwg}, the ``complete'' spectra of fully-heavy $4c$, $4b$, $bc \bar b \bar c$ and $bb\bar c \bar c$ compact tetraquarks were computed in the relativized diquark model.
Moreover, by making use of the baryon-meson (or quark-diquark) supersymmetry, the masses of the $3b$ and $3c$ baryons were calculated in both the  Godfrey and Isgur's relativized QM and the relativized quark diquark model. These predictions for the masses of $3b$ and $3c$ baryons are similar to those of lattice QCD calculations.

Ref. \cite{Anwar:2017toa} was cited twice by LHCb: firstly, in the paper on the search for the fully-bottom tetraquark decaying into four muons; secondly, in the paper on the discovery of the $X(6900)$ tetraquark.
Ref. \cite{Bedolla:2019zwg} was cited in the $X(6900)$ tetraquark discovery paper by LHCb.
\\

In Ref. \cite{Becchi:2020uvq}, production cross-sections and branching ratios for fully-charmed $0^{++}$ and $2^{++}$ tetraquarks were calculated, and it was pointed out that LHCb could observe a fully-charm tetraquark with the present luminosity. Moreover, it was shown that the $2^{++}$ was more likely to be observed because of the higher production cross-section and the higher branching ratio in the di-$J/\Psi$ channel. 
After this paper had been uploaded on the arXiv, LHCb uploaded its results on the $X(6900)$ fully-$c$ tetraquark. It should be noted that the experimental width of the $X(6900)$, which is $80\pm19\pm33$ MeV, is compatible with the predictions of Ref. \cite{Becchi:2020uvq} within the experimental error.

\hfill\break

{\bf   Coupled-channel model for heavy quarkonium-like mesons and the $X(3872)$ as a core + four quark $q \bar c - c \bar q$ components at threshold.} In Ref. \cite{Ferretti:2018tco}, a coupled-channel model (CCM) for heavy quarkonium-like mesons was developed in which the quarkonium-like mesons are described as a $Q \bar Q$ core, together with $q \bar Q - Q \bar q$ components. 
This CCM was used to study the masses of $\chi_{\rm c}(2P)$ and $\chi_{\rm b}(3P)$ states with threshold corrections. It was found that the $\chi_{\rm c}(2P)$ states, and the $X(3872)$ in particular, could contain non-negligible  threshold components, while the $\chi_{\rm b}(3P)$ bottomonia are expected to show very small threshold components \cite{Ferretti:2018tco}. The 
calculated $\chi_{\rm c}(2P)$ spectrum is in good agreement with the experimental data.

The hidden-flavor $J/\psi \rho$ and $J/\psi \omega$ transitions of the $X(3872)$ have also been computed, by combining the CCM formalism with a diagrammatic non-relativistic approach to meson-meson scattering.
The result \cite{Ferretti:2018tco}
\begin{equation}
R_{\omega/\rho} \equiv \frac{\Gamma\!\left(X(3872) \to J/\psi \omega \right)}{\Gamma\!\left(X(3872) \to J/\psi \rho \right)} = 0.6
\end{equation}
is compatible with the current experimental data, $R_{\omega/\rho} = 0.8\pm0.3$\, 
, within the experimental error.

%% file: rosner.tex
\vfill\eject

\section{%
Exotics, \boldmath $QQq$ baryons, tetraquarks and pentaquarks}
\centerline{\bf\large Marek Karliner$^{1*}$ and Jonathan L. Rosner$^{2\dagger}$}
\centerline{\vrule width 0pt height 3ex
$^1$School of Physics and Astronomy, Tel Aviv University, Tel Aviv 69978, Israel}
\centerline{$^2$Enrico Fermi Institute and Department of Physics}
\centerline{University of Chicago, 5640 S. Ellis Avenue, Chicago, IL 60637, USA}
\centerline{\tt $^*$marek@tauex.tau.ac.il \quad $^\dagger$rosner@hep.uchicago.edu}
\hfill\break

\noindent
The following is a list of key references in our work on these subjects. 

\bigskip

{\bf Prediction of exotic mesons from $s$-channel--$t$-channel duality:}

J. Rosner, Possibility of baryon-antibaryon enhancements
with unusual quantum numbers \cite{Rosner:1968si}.  In
baryon-antibaryon scattering, $qqq$-$\bar q \bar q \bar q$
$\to$ a qqq-$\bar q \bar q \bar q$, a non-exotic $t$-channel
($q \bar q$) exchange is dual to an exotic
$q q \bar q \bar q$ state in the $s$ channel.  
\smallskip

\includegraphics[width=0.9\textwidth]{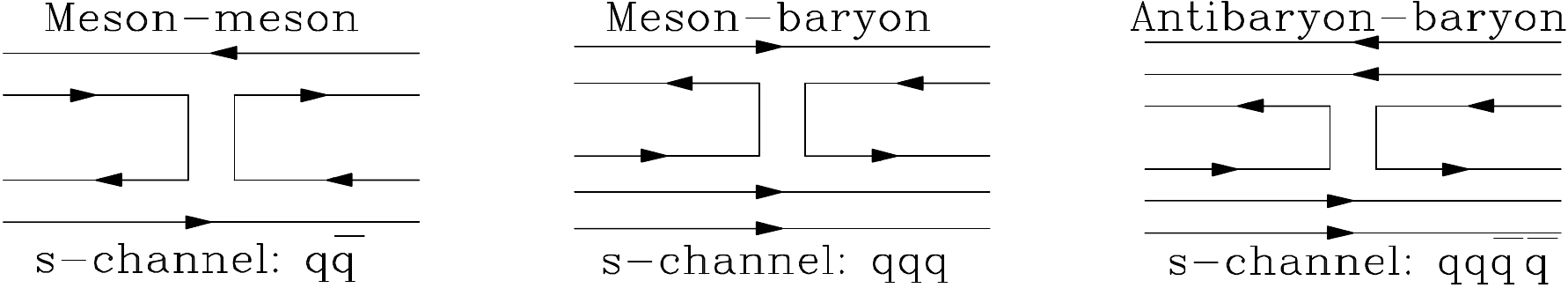}

\medskip

{\bf Successful prediction of bottom baryon masses:}

The quark model and $b$ baryons, Marek Karliner, Boaz Keren-Zur, Harry J.
Lipkin, and Jonathan L. Rosner \cite{Karliner:2008sv}.  Contributions of
constituent masses, color hyperfine interactions, and phenomenological
interquark potentials reproduce masses of states such as $\Xi_b^- =
bsd,~\Xi_b^0 = bsu$, and $\Omega_b^- = bss$. This calibration of spin-dependent
interaction between quarks and calibration of the extra attraction between
eavier quarks have been essential for exotics.

\medskip

{\bf Successful prediction of $\Xi_{cc}^{++}$ doubly charmed baryon mass:}

Baryons with two heavy quarks:  Masses, production, decays, and
detection, Marek Karliner and Jonathan L. Rosner \cite{Karliner:2014gca}.
Previously validated methods and an {\it ansatz} relating the effective
mass of the $cc$ color antitriplet diquark to that of the color-singlet
quarkonium mass allow the prediction of the mass of the $\Xi_{cc}^{++}
= ccu$ state to within several MeV of its value subseqently observed by
the LHCb detector.  A crucial step in predicting $M(T_{cc}^+)$ (see below).

\medskip

{\bf Successful prediction of pentaquark masses:}

New exotic meson and baryon resonances from doubly-heavy hadronic
molecules, Marek Karliner and Jonathan L. Rosner \cite{Karliner:2015ina}.
In the decay $\Lambda_b \to J/\psi p K^-$, resonant activity in the $J/\psi p
= c \bar c u u d$ channel is anticipated and observed at the threshold masses
of $\Sigma_c \bar D^*$ and $\Sigma_c \bar D$.

\medskip

{\bf Successful prediction of mass of tetraquark $T_{cc} = c c \bar u \bar d$:}

Discovery of doubly-charmed $\Xi_{cc}$ baryon implies a stable $bb \bar u
\bar d$ tetraquark, Marek Karliner and Jonathan L. Rosner
\cite{Karliner:2017qjm}.  Use is made of previously mentioned ansatz for
masses of heavy-quark color antitriplet to anticipate masses of $T_{cc},~
T_{cb}$, and $T_{bb}$.  LHCb has observed a candiate for $T_{cc}$ within
several MeV of its predicted mass.  First robust prediction of a $T_{bb}$
tetraquark stable against strong decay.

\medskip

{\bf Successful prediction of mass of first exotic hadron with open heavy
flavor:}

First exotic hadron with open heavy flavor: $cs\bar u\bar d$ tetraquark,
Marek Karliner and Jonathan L. Rosner \cite{Karliner:2020vsi}.  Evidence for
a string-junction picture of non-exotic and exotic mesons and baryons is
presented (see figure), and a peak in the $D^+ K^-$ channel is anticipated.
\smallskip

 \hskip 0.5in \includegraphics[width=0.45\textwidth]{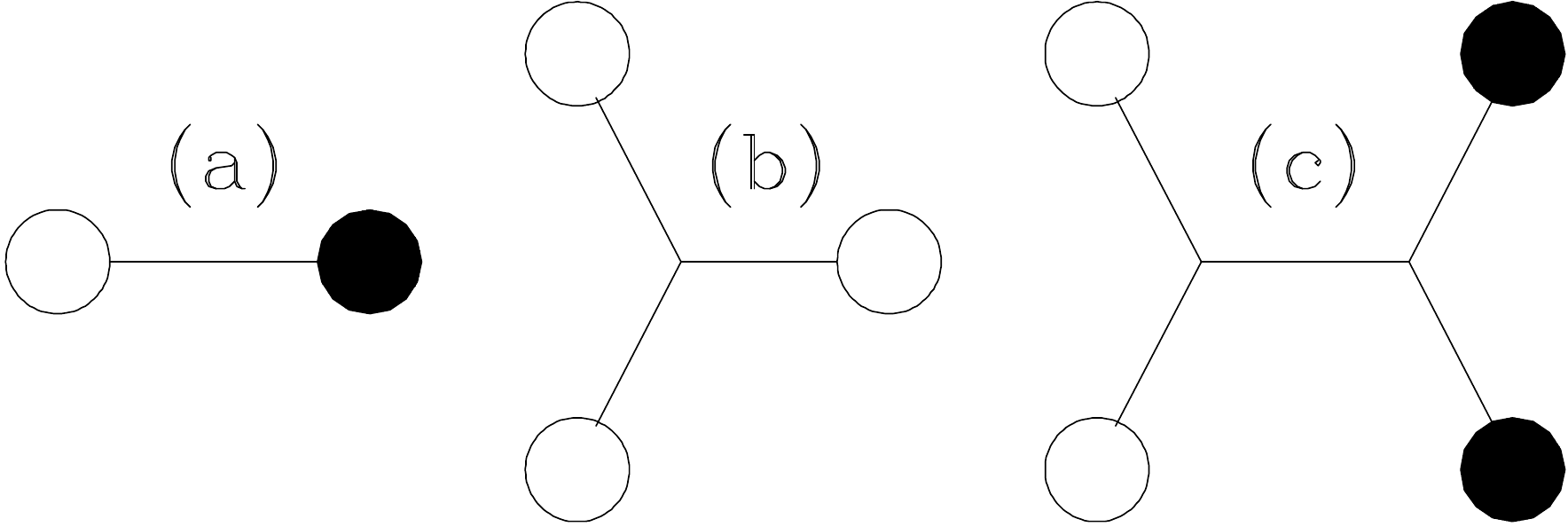}
\vskip -0.9in
\parindent 4.5in
(a) Non-exotic meson

(b) Non-exotic baryon

(c) Exotic meson

\parindent 0in

\vfill\eject

%% file: hosaka.tex
\vfill\eject
\section{Exotic structures of doubly-heavy tetraquarks}
\centerline{\large\bf Atsushi Hosaka}
\centerline{Research Center for Nuclear Physics (RCNP)}
\centerline{Osaka University, Ibaraki, 567-0047, Japan}
\centerline{\tt hosaka@rcnp.osaka-u.ac.jp}
\hfill\break

We have studied the exotic structures of doubly-heavy tetraquarks 
$QQ\bar q \bar q$ for both bound and resonant states.  

In \cite{Meng:2020knc}, stable doubly-heavy tetraquarks are investigated.  
The quark model hamiltonian is rigorously diagonalized for four-body states by the Gaussian expansion method.  
It has been confirmed that doubly-bottomed tetraquarks may accommodate 
a deeply bound state of $J^P = 1^+$ with a binding energy 
of order 100 MeV or more.
In addition, we have found a shallow bound state, the existence of which 
depends sensitively on the strength of attraction, though. 
The deep and shallow states develop qualitatively 
different internal structure: 
the deeply-bound one is like a three-body state of 
doubly heavy diquark $[bb]$ and 
two light quarks in the good-diquark channel, 
while the shallow one looks like a $BB^*$ molecule, 
reflecting a general feature that states near a threshold 
develop hadronic molecules.  
\par
Furthermore, in Ref.~\cite{Meng:2021yjr}, resonant states are studied
by including couplings to meson-meson ($Q\bar q$-$Q\bar q$) scattering
states in the four-body calculation.  
It was found that heavy quark triplet of $J^P = 0^-, 1^-, 2^-$ may exist as resonances.  

%% file: brambilla.tex
\vfill\eject
\color{black}
\section{Effective Field Theories for \boldmath the $X$, $Y$, $Z$ frontier}
\centerline{\bf \large Nora Brambilla}
\centerline{Physik Department, Technische Universität München,}
\centerline{James-Franck-Strasse 1, 85748 Garching, Germany}
\centerline{Institute for Advanced Study, Technische Universität München,}
\centerline{Lichtenbergstrasse 2a, 85748 Garching, Germany}
\centerline{Munich Data Science Institute, Technische Universität München,}
\centerline{Walther-von-Dyck-Strasse 10, 85748 Garching, Germany}
\centerline{\tt nora.brambilla@ph.tum.de}
\hfill\break

\def \m2   {\mu^{2 \epsilon}}
\def\lQ{\Lambda_{\rm QCD}}
\def\als{\alpha_{\rm s}} 
\def\bea{\begin{eqnarray}}
\def\eea{\end{eqnarray}}
\def\be{\begin{equation}}
\def\ee{\end{equation}}
\def\a{\alpha}  \def\g{\gamma} \def\G{\Gamma}
\def\la{\lambda} \def\lap{\lambda^{\prime}} \def\pa{\partial} \def\de{\delta} \def\De{\Delta} \def\dag{\dagger}
\def\e{\epsilon} \def\nb{\bm{\nabla}}  \def\Oc{{\rm O}} \def\S{{\rm S}}
\def\bnabla{{\bm \nabla}}                                                                
\def\bsigma{{\bm \sigma}}


Exotic states have been predicted before and after the advent of QCD.
In the last decades they have been observed at accelerator experiments  in
the sector with two heavy quarks, at or above the quarkonium strong decay
threshold. They are called $X$, $Y$, $Z$ states.
These states  offer a unique possibility for investigating the dynamical
properties of strongly correlated systems in QCD.
I report here  how an alliance of nonrelativistic effective field theories and
lattice can allow us to address these states in QCD. 


The $X$, $Y$, $Z$ states have been discovered  in the sector  with a heavy quark and an antiquark, or two heavy quarks, 
 at or above the strong decay threshold. The theory of $Q \bar{Q}$ states below  threshold has 
 been constructed in the last decades and it is based on the nonrelativistic effective field theory 
 called potential Nonrelativistic QCD (pNRQCD) \cite{Brambilla:2019esw}.
 It allows to systematically define and calculate the potentials 
 and provides a scheme to calculate quarkonium observables. 
 
 The theory is constructed to be equivalent to QCD 
 and is endowed with a power-counting that allows to attach  errors to physical predictions.
 It is based on scale factorizations  and allows to factorize nonperturbative low energy contributions inside  gauge invariant correlators that do not depend 
 on flavor. Such correlators may be calculated on the lattice or extracted from the data. This greatly  boosts predictivity and allows for model-independent predictions, 
 which have been recently extended to address quarkonium production 
 and the nonequilibrium evolution of quarkonium in medium.

In the most interesting region,  close or above the strong decay  threshold, where the $X$, $Y$, $Z$ 
have been discovered, the situation is much more complicated: 
 there is no mass gap between quarkonium and the creation of a couple of heavy-light mesons, nor to gluon 
 excitations. Therefore  many additional states with the light-quark quantum numbers may appear.
 Still, $m$  is a  large scale and   a first-scale factorization is applicable, so that
 nonrelativistic QCD is still valid.
 Then, if we want to introduce a description   of the bound state similar to pNRQCD,
 making apparent that the zero-order problem is the Schr\"odinger equation,
 we can still count on  another scale separation.
 
Let us consider   bound states of two nonrelativistic particles and some light d.o.f., 
e.g. molecules in QED  or quarkonium  hybrids ($Q\bar{Q} g$ states) 
or tetraquarks  ($Q\bar{Q}   q\bar{q}$  states) in QCD:
electron/gluon fields/light quarks  change adiabatically in the presence of heavy quarks/nuclei.
  The heavy quarks/nuclei interaction may be described at leading 
order in the nonrelativistic expansion by an effective  potential $V_\kappa$ between the static sources, where $\kappa$ labels different excitations 
of the light degrees of freedom. 
 A plethora of states  can be  built on each on the potentials $V_\kappa$,  by solving the
corresponding Schr\"odinger equation.  
This picture corresponds to the Born-Oppenheimer (BO) approximation. 
Starting from pNRQED/pNRQCD the BO approximation can be made 
rigorous and cast into a suitable EFT called Born-Oppenheimer EFT (BOEFT)
\cite{Berwein:2015vca,Brambilla:2017uyf,Brambilla:2018pyn,Brambilla:2019esw}
which exploits the hierarchy of scales
$\lQ \gg mv^2$, $v$ being the velocity of the heavy quark.

\begin{figure}[ht]
 \centering
  \begin{minipage}[b]{0.9\textwidth}
    \includegraphics[width=\textwidth]{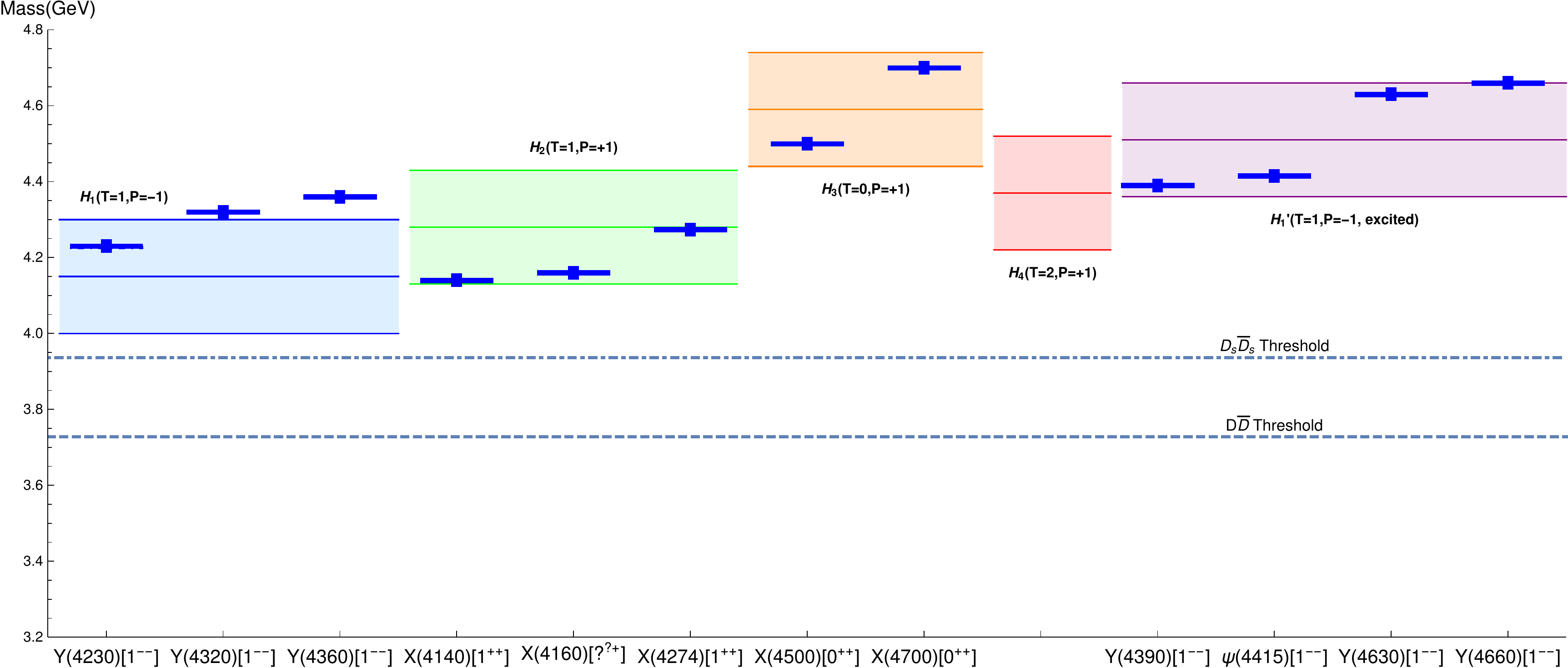}  
    \caption{Mass spectrum of neutral exotic charmonium states obtained by solving the BOEFT coupled Schr\"odinger equations.
     The neutral experimental states  that have matching quantum numbers are plotted in solid blue lines. In the figure $T$ stay for the total angular momentum. $H_1^\prime$ is the first $H_1$ radial excitation of $H_1$.
     The multiplets have been plotted with error bands corresponding to a gluelump mass uncertainty of 0.15 GeV. Figure taken from
     \cite{Brambilla:2019esw}.} \label{figexp}
     \end{minipage}
\end{figure}

 In  \cite{Berwein:2015vca} we have obtained the BOEFT that describes hybrids. In particular, 
 we have obtained the static potentials and the  set of coupled 
 Schr\"odinger equations, solved them and produced all the hybrids multiplets, see Fig. \ref{figexp}.
 We observed  a  phenomenon called $\Lambda$ doubling, known in
 molecular physics, but with smaller size. This and the structure of the multiplets differ 
 from what is obtained in models cf.  \cite{Berwein:2015vca}.
 We used lattice input on the hybrid static energies and on the gluelump mass.
 
  In \cite{Brambilla:2018pyn}  we obtained the spin-dependent potentials at order $1/m$ and $1/m^2$ 
  in the quark mass expansion and thereby we could calculate all the hybrids spin multiplets. 
Notice that on one hand one seldom finds  spin interaction considered in models, 
on the other hand it would be different.
In fact,  the ${\cal O}(1/m)$ contributions couple the angular momentum of the gluonic excitation 
with the total spin of the heavy-quark-antiquark pair.
These operators are characteristic of the hybrid states and are absent in standard quarkonia. 
Among the ${\cal O}(1/m^2)$ operators,  besides the standard spin-orbit, total spin squared, and tensor spin operators  which appear for standard quarkonia,
three novel operators appear. 
So interestingly, differently from the quarkonium case, the hybrid potential gets a first contribution 
already at order $\Lambda^2_{\text{QCD}}/m$.
Hence, spin splittings are remarkably less suppressed in heavy quarkonium hybrids than in heavy quarkonia: 
this will have a notable impact on the phenomenology of exotics.

We extracted the nonperturbative low-energy correlators appearing in the factorization, 
fixing them from lattice data on 
of charmonium hybrids. We could then predict all spin multiplets of the bottomonium hybrids,
which are significantly more difficult to evaluate on the lattice.
In the same framework it is also possible to calculate hybrids' decays and quarkonium/hybrids mixing.

 The BOEFT may also be used to describe tetraquarks, double heavy baryons and pentaquarks \cite{Brambilla:2017uyf}.
 In the case of tetraquarks,  a necessary input is the calculation of the 
generalized Wilson loops with appropriate symmetry and light quark operators on the lattice, 
so that besides the quantum number
$\kappa$ also the isospin quantum numbers  $I=0, 1$ have to be considered.

The BOEFT approach reconciles  the different pictures of exotics based on tetraquarks, molecules, 
hadroquarkonia\ldots\ 
In fact, for a $Q\bar{Q} q\bar{q} $ or $Q\bar{Q}g$ state static energy exhibits different regimes 
as a function of distance scale: 
a hadroquarkonium picture for short distances, then a tetraquark (or hybrid), and after crossing the 
heavy-light meson line, threshold effects need to be taken into account and a molecular picture emerges.
QCD dictates, through lattice correlators and the BOEFT characteristics and power-counting, 
which structure dominates and in which precise way.
In addition to the above discussion,  production and suppression in medium may be described in the same approach.

%% file: yamaguchi.tex
\vfill\eject

\section{\boldmath  Pentaquarks as a mixture of the compact five-quark states and hadronic molecules}  
\centerline{\bf\large Yasuhiro Yamaguchi}
\centerline{Adv. Science Research Center, Japan Atomic Energy Agency (JAEA), Tokai 319-1195, Japan}
\centerline{\tt yamaguti@rcnp.osaka-u.ac.jp}
\hfill\break

The heavy quark spin symmetry (HQSS) is one of the important attributes of heavy hadron systems, 
including exotics. 
Ref.~\cite{Hosaka:2016ypm} 
discussed
the mass degeneracy of hadron composite systems containing one heavy quark, and the one pion exchange 
potential (OPEP), enhanced by the HQSS.
The features of the HQSS obtained in these studies will be helpful in understanding heavy exotics.

\begin{figure}[htbp]
 \begin{center}
  \includegraphics[width=0.64\linewidth,clip]{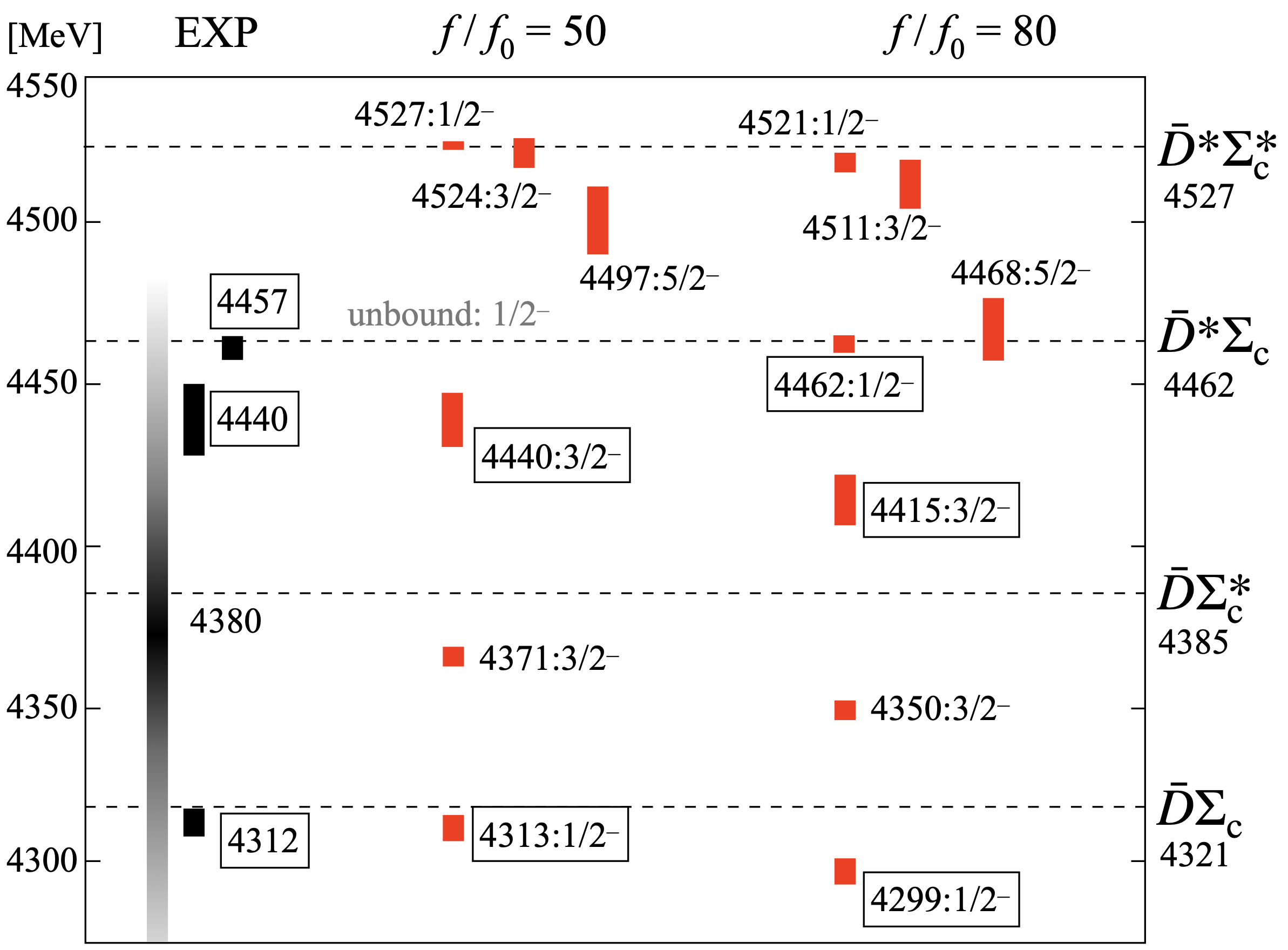}
  \caption{\label{fig:pentaquarks} LHCb data (EXP) and obtained masses and widths for $P_c$ pentaquarks, taken from Ref.~\cite{Yamaguchi:2019seo}.
  The centers of the bars are located at the pentaquark masses, while their lengths corresponds to the decay widths.
  The horizontal dashed lines show the $\Sigma_c^{(\ast)}\bar{D}^{(\ast)}$ thresholds. }
 \end{center}
\end{figure}

The $P_c$ pentaquarks discovered by the LHCb experiment have attracted much interest because their decay products
indicate they are five-quark states.  It is not yet understood, however, how the five quarks are arranged inside the
$P_c$. We have discussed a possible exotic structure of $P_c$,  as
a mixture of a compact five-quark state and hadronic molecules involving a 
$\bar{D}^{(\ast)}$ meson and $Y_c = \Lambda_c,\Sigma_c^{(\ast)}$ baryon~\cite{Yamaguchi:2019seo}.
The model accurately reproduced the observed mass and width of the $P_c$ states,
as shown in Fig.~\ref{fig:pentaquarks}.
The roles of the hadron interactions in the $P_c$ resonances were also discussed.

In addition, studies of the OPEP and the role of the tensor interaction have been performed  
in Ref.~\cite{Yamaguchi:2019vea}, yielding properties of selected exotics.

